\documentclass{jfm}

\usepackage{graphicx}
\usepackage{natbib}

\ifCUPmtlplainloaded \else
  \checkfont{eurm10}
  \iffontfound
    \IfFileExists{upmath.sty}
      {\typeout{^^JFound AMS Euler Roman fonts on the system,
                   using the 'upmath' package.^^J}%
       \usepackage{upmath}}
      {\typeout{^^JFound AMS Euler Roman fonts on the system, but you
                   dont seem to have the}%
       \typeout{'upmath' package installed. JFM.cls can take advantage
                 of these fonts,^^Jif you use 'upmath' package.^^J}%
      }
  \else
  \fi
\fi

\ifCUPmtlplainloaded \else
  \checkfont{msam10}
  \iffontfound
    \IfFileExists{amssymb.sty}
      {\typeout{^^JFound AMS Symbol fonts on the system, using the
                'amssymb' package.^^J}%
       \usepackage{amssymb}%
       \let\le=\leqslant  
         
      }{}
  \fi
\fi

\ifCUPmtlplainloaded \else
  \IfFileExists{amsbsy.sty}
    {\typeout{^^JFound the 'amsbsy' package on the system, using it.^^J}%
     \usepackage{amsbsy}}
    {\providecommand\boldsymbol[1]{\mbox{\boldmath $##1$}}}
\fi

\providecommand\bnabla{\boldsymbol{\nabla}}

\newsavebox{\astrutbox}
\sbox{\astrutbox}{\rule[-5pt]{0pt}{20pt}}

\newcommand\etal{\mbox{\textit{et al.}}}

\title[Rotating homogeneous turbulence]{Structure and decay of rotating homogeneous turbulence}

\author[M. Thiele, W.-C. M\"uller]%
{M\ls A\ls R\ls K\ns T\ls H\ls I\ls E\ls L\ls E$^1$%
\ns \and W\ls O\ls L\ls F\ls -\ls C\ls H\ls R\ls I\ls S\ls T\ls I\ls A\ls N \ns M\ls \"U\ls L\ls L\ls E\ls R$^{2}$}

\affiliation{$^1$Physikalisches Institut, Universit\"at Bayreuth, D-95440 Bayreuth, Germany\\[\affilskip]
$^2$Max-Planck-Institut f\"ur Plasmaphysik, D-85748 Garching, Germany}

\date{?? and in revised form ??}

\begin{document}

\maketitle

\begin{abstract}
Navier-Stokes turbulence subject to solid-body rotation is studied by
high-resolution direct numerical simulations (DNS) of freely decaying
and stationary flows.  Setups characterized by different Rossby
numbers are considered.  In agreement with experimental results strong
rotation is found to lead to anisotropy of the direct nonlinear
energy flux, which is attenuated primarily in the direction of the rotation axis.  In
decaying turbulence the evolution of kinetic energy follows an 
approximate power law with a distinct dependence of the
decay exponent on the rotation frequency. A simple phenomenological
relation between exponent and rotation rate reproduces this
observation.  Stationary turbulence driven by large-scale forcing
exhibits $k_\perp^{-2}$-scaling in the rotation-dominated inertial range of
the one-dimensional energy spectrum taken perpendicularly to the rotation axis.
The self-similar scaling is shown to be the cumulative result of individual spectral contributions which, for low rotation rate, display $k_\perp^{-3}$-scaling near the $k_\parallel=0$ plane. 
A phenomenology which incorporates the
modification of the energy cascade by rotation is proposed. In the observed 
regime the nonlinear turbulent interactions are strongly influenced by rotation but 
not suppressed.  
{Longitudinal two-point velocity structure functions taken perpendicularly to 
the axis of
rotation indicate weak intermittency of the $k_\parallel=0$ (2D) component of the 
flow while the intermittent scaling of $k_\parallel\neq 0$ (3D) fluctuations is well
captured by a modified She-L\'ev\^eque intermittency model which yields the expression 
$\zeta_p=p/6 + 2(1-(2/3)^{p/2})$ for the structure function scaling exponents.
}
\end{abstract}

\section{Introduction}\label{sec:intro}
Hydrodynamic turbulence subject to rotation is an ubiquitous problem in 
fluid mechanics. The understanding of its detailed properties is crucial for
engineering problems such as the design of turbomachinery and for the understanding
of atmospheric and oceanic flows influenced by the earth's rotation. 
These applications have motivated an extensive research on the macroscopic
and spectral properties of rotating turbulence though a comprehensive and fully
consistent physical picture is still missing. 

Early experiments show the tendency of rotating turbulence to
asymptotically two-dimensionalize in planes perpendicular to the
rotation axis $\boldsymbol{\Omega}$, where $\Omega= |\boldsymbol{\Omega}|$ is the rotation frequency. Signatures of this behaviour
are, e.g., an increased ratio of velocity correlation lengths along and perpendicular to 
the axis of
rotation, \cite[see, for example,][]{ibbetson_tritton:exp}, visible elongation of spatial
structures along $\boldsymbol{\Omega}$,
\cite[e.g. in][]{hopfinger_browand_gagne:exp}, or anisotropy of characteristic
length- and time-scales \cite[e.g.][]{wigeland:exp,jacquin_etal:exp}.
In addition, slower decay of kinetic energy is found in the works of
\cite{wigeland:exp} and \cite {jacquin_etal:exp} as compared to
non-rotating turbulence.  More recently, the experiments of
\cite{morize_moisy:exp} show the significant influence of confinement
on the decay of rotating turbulence.
\cite{baroud_etal:expspec,baroud_etal:expstruc} consider two-point
increment statistics up to order $p=10$ yielding inertial-range
scaling exponents $\zeta_p=p/2$. The work of
\cite{morize_moisy_rabaud:exp} focuses on the energy spectrum $E(k)$ which
exhibits a Rossby number dependent scaling exponent continuously
ranging between -5/3 (slow rotation) and about -2.3 (fast rotation). 
Both
experiments feature turbulent flows with energy injection at
intermediate scales and thus allow an inverse cascade of kinetic energy to 
develop in the quasi-two-dimensional flow expected for strong rotation. 

The first direct numerical simulations of rotating turbulence
conducted by \cite{bardina_ferziger_rogallo:rotdns32},
\cite{mansour_cambon_speziale:rotdns32} and \cite{hossain:rotdns32}
also suggest reduction of the energy flux and
two-dimensionalization. They suffer, however, from small spatial
resolution and correspondingly low Reynolds numbers  and 
do not yield conclusive results on the spectral
properties of the flow.
Simulations at higher Reynolds number by \cite{godeferd_lollini:rotdns48} and
\cite{morinishi_nakabayashi_ren:rotdns128} provide more evidence for
the experimentally observed behaviour of the correlation lengths and
the decay properties of the kinetic turbulent energy,
respectively. For the case of driven turbulence
\cite{yeung_zhou:rotdns256} find $k^{-2}$-scaling of the energy
spectrum in a setup with large scale forcing. On the contrary, the
simulations of \cite{smith_waleffe:rotdns200},
\cite{chen_chen_eyink_holm:rotdns128} and \cite{mininni_etal:scale_interactions} explore systems with forcing at
intermediate scales. All three works identify a quasi-two-dimensional state
in the case of strong rotation accompanied by an inverse energy
cascade for wavenumbers $k<k_{\mathrm{forcing}}$. In this
wavenumber range \cite{smith_waleffe:rotdns200} report to find $k^{-3}$-scaling of the energy spectrum while a $k^{-2}$-behaviour for 
$k>k_\mathrm{forcing}$ is observed by \cite{smith_waleffe:rotdns200} and \cite{mininni_etal:scale_interactions}.

Large-Eddy simulations like those of \cite{squires_etal:rotles128512},
\cite{bartello_metais_lesieur:rotles64} and
\cite{yang_domaradzki:rotles64} are a useful tool to study the effects
of rotation on characteristic properties of rotating turbulence such
as the slowing down of energy decay, the large-scale structure of the
flow and its tendency to become two-dimensional. The approach allows
to attain Reynolds numbers beyond the reach of DNS.  However, the
results of LES crucially depend on the applied subgrid model which
requires careful adjustments and gauging by comparison with
high-resolution DNS or experimental measurements. For the detailed
numerical investigation of small-scale properties of turbulence like
anisotropic inertial-range scaling of the energy spectrum and
higher-order structure functions, therefore, the DNS approach is
chosen in this work.

Different theoretical approaches exist with regard to rotating
turbulence. Shell-models, which are strongly simplified representations of nonlinear turbulent
 dynamics,
allow for the straightforward inclusion of rotation effects, e.g. 
via modes governed by a time-correlated stochastic process as in 
\cite{hattori_rubinstein_ishizawa:rotshell}. The simplicity of those models, however, restricts
their predictive capabilities to a continuous change of spectral scaling between Kolmogorov 
$k^{-5/3}$-scaling (slow rotation) and $k^{-2}$-behaviour for fast rotation.
\cite{zhou:dimanal} and \cite{mahalov_zhou:dimanal}
recover $k^{-2}$-scaling of the energy spectrum in the context of
quasi-normal closure theory using dimensional arguments first employed
by \cite{kraichnan:ikmodel}. \cite{canuto_dubovikov:formal} arrive at
similar results via a formal treatment of the spectral energy flux
using helical-mode decomposition.  The former work is based on the
assumption that $\tau_\Omega=\Omega^{-1}$ is the dominant time scale of nonlinear relaxation
while the latter study is analytically showing the same covering cascade dynamics
for rotation rates up to $\tau_\Omega\simeq \tau_\mathrm{NL}\sim \ell/v_\ell $.
Here, $\tau_\mathrm{NL}$ is the
nonlinear eddy-turnover or relaxation time defined with the RMS
velocity $v_\ell$ at scale $\ell$ (see below). 
On the contrary, weak turbulence theory 
\cite[see, e.g.,][]{cambon_rubinstein_godeferd:rotweak,galtier:rotweak,bellet_etal:weaksimul} 
yields different anisotropic scaling relations
but is only strictly applicable in the asymptotic limit 
$\tau_\Omega\ll\tau_\mathrm{NL}$.

This article is motivated by the lack of a simple dynamical picture of
spectral energy transfer which could also help
to improve the understanding of turbulent energy decay. The proposed
phenomenology is backed-up by high-resolution direct numerical
simulations of incompressible rotating homogeneous turbulence in free
decay and of rotating turbulence subject to large-scale driving.  The
kinetic energy is found to exhibit approximate power-law decay with an
exponent that decreases with increasing rotation frequency. This
behaviour is in agreement with the observed attenuation of nonlinear
spectral transfer under the influence of rotation and is captured by a
simple model based on the cascade phenomenology.  In the case of
forced turbulence, analysis of one-dimensional spectral data indicates
anisotropic energy flux mainly perpendicular to
$\boldsymbol{\Omega}$. Inertial-range scaling of the perpendicular energy spectrum
$\sim k_\perp^{-2}$ is observed and reproduced by the dynamical
cascade phenomenology.  While the parallel spectrum 
does not show clear
scaling, the perpendicular spectra at fixed $k_\parallel$ 
display $k^{-3}$-scaling in the vicinity of $k_\parallel=0$ for low rotation rate. For growing $\Omega$, energy accumulates 
at $k_\parallel=0$ in agreement with the observed quasi-two-dimensionalization of the system.
{The intermittency measured via velocity structure functions
perpendicular to $\boldsymbol{\Omega}$ is very weak for the 2D component of the flow  
($k_\parallel=0$). In contrast, the intermittent scaling signature of the 3D component 
($k_\parallel\neq 0$) is in agreement with a modified She-L\'ev\^eque model and exhibits a 
weak trend towards Gaussianity for high rotation rate.}
The remainder of this paper is organized in the
following way: In section 2 we introduce the equations of rotating
fluid flow and present the numerical methods that have been
used. Sections 3 and 4 comprise the results on spectral and
macroscopic properties of the flow and a summary is given in section
5.

\section{Basic equations and numerical methods}\label{sec:model}
Incompressible fluid flow subject to solid-body rotation is usually
described in a frame of reference rotating about a fixed axis
$\boldsymbol{\Omega}=\Omega\boldsymbol{\hat{e}}_z$ with frequency $\Omega$ using the dimensionless Navier-Stokes equations including
the Coriolis force,
\begin{eqnarray}
  \partial_t\boldsymbol{v} + \boldsymbol{v}\cdot\nabla\boldsymbol{v} &=& -\nabla P +\mu \Delta\boldsymbol{v} + 2\,{\Omega}\,\boldsymbol{v} \times \boldsymbol{\hat{e}}_z,\label{eq:nast}\\
  \nabla\cdot\boldsymbol{v} &=& 0\,,
\end{eqnarray}
where the centrifugal force has been incorporated in the generalized
pressure $P$, or the numerically more favourable vorticity formulation
\begin{eqnarray}
  \partial_t\boldsymbol{\omega} &=& \nabla \times (\boldsymbol{v} \times \boldsymbol{\omega} + 2\,{\Omega}\,\boldsymbol{v} \times \boldsymbol{\hat{e}}_z ) + {\mu} \Delta\boldsymbol{\omega},\label{eq:vort}\\
  \nabla\cdot\boldsymbol{v} &=& 0\,\label{eq:incomp}
\end{eqnarray}
with the vorticity $\boldsymbol{\omega} = \nabla \times
\boldsymbol{v}$ \cite[see, e.g.,][]{greenspan:book}. The dimensionless kinematic viscosity ${\mu}$ and rotation rate
${\Omega}$ are both assumed to be constant.

Equations (\ref{eq:vort}) and (\ref{eq:incomp}) are integrated
numerically on a cubic box extending $2\pi$ in each dimension with
triply periodic boundary conditions. The integration is
performed in a pseudospectral Fourier representation by applying an explicit
leapfrog scheme. The diffusive term is incorporated using an
integrating factor technique \cite[see, e.g.,][]{meneguzzi_pouquet:convecdynamo}.
The time step which is required for numerical stability by the Courant-Friedrichs-Lewy criterion 
ensures the temporal resolution of all inertial-wave oscillations present
in the system. 
The aliasing error introduced by the pseudospectral approach is reduced by
spherical mode truncation. The quasi-stationary state of the driven systems
is sustained by a forcing which freezes
all modes with $k \le k_f = 2$. The applied resolution is $512^3$
collocation points for forced and $256^3$ for decaying turbulence.

The isotropic initial state of the forced simulations is a smooth
vorticity field with random phases and an energy spectrum $E(k) \sim
\exp\left(-k^2/k_0^2\right)$, $k_0=4$. This configuration
is integrated forward in time without forcing until the maximum
of enstrophy is reached. After this period of time, which corresponds to one
large-eddy-turnover time (LET) and to three units of time in the notation of this paper, 
the energy spectrum is fully developed in its spectral extent and the forcing is switched on.
The spectral distribution of energy evolves self-consistently towards a state with 
a low-order power-law at largest scales and a Kolmogorov-type inertial range.   
As soon as total energy $E = \int_V dV v^2/2$ and dissipation $\varepsilon = {\mu} \int_V
dV \omega^2$ are statistically stationary with $E\simeq 1$ (more precisely: $0.87$ for $\Omega=5$ and
$1.3$ for $\Omega=50$) and
$\varepsilon\simeq 0.05$, ${\Omega}$ is set to a finite value.  
For decaying turbulence simulations a similar random initial
field is generated and left to evolve freely for one LET before rotation is switched on.
After the onset of rotation a statistically stationary state of
turbulence is maintained in driven simulations for 10 or 15 large-eddy turnover
times depending on the rotation rate. Decaying runs extend over approximately 12 LETs.  
The dimensionless kinematic viscosity is constant in all simulations. 

The characteristic length $L_0$
and velocity $V_0$, necessary for the calculation of the macroscopic
Rossby number, $\mathsf{Ro}=V_0/(2\Omega L_0)$, and Reynolds number,
$\mathsf{Re}=L_0V_0/\mu$, can only be determined a posteriori in
homogeneous turbulence. Therefore, both quantities are estimated in the dimensionless system
using $E$, $\varepsilon$, and $\Omega$ as $L_0\sim
E/({\Omega}\varepsilon)^{1/2}$ and $V_0\sim E^{1/2}$. Hence Rossby
and Reynolds numbers given in this paper are defined as
$\mathsf{Ro}=\sqrt{\varepsilon/(4{\Omega} E)}$ and
$\mathsf{Re}=\sqrt{E^3/({\Omega}\varepsilon)}/{\mu}$, respectively.
Specific parameters of the performed simulations are summarized in
table \ref{tab:parameters}.
\begin{table}
  \begin{center}
  \begin{tabular}{llllllll}
      simulation   & grid & forcing & ${\Omega}$   &   ${\mu}$ &$t_\mathrm{Sim}$&\textsf{Ro}&\textsf{Re}\\[3pt]
       I   & $512^3$ &yes& 5 & $4\times10^{-4}$&45&$5.3\times 10^{-2}$&4000 \\
       II  & $512^3$ &yes& 50 & $4\times10^{-4}$&30&$1.3\times 10^{-2}$&2300 \\ 
       III & $256^3$ &no& 0-5 & $1\times10^{-3}$&36& $\infty$ -- $2\times 10^{-2}$ & 300--1100\\
  \end{tabular}
  \caption{Parameters for the different simulation runs: rotation rate $\Omega$, 
           viscosity $\mu$, total duration $t_\mathrm{Sim}$, Rossby number \textsf{Ro}, and
           Reynolds number \textsf{Re}}
  \label{tab:parameters}
  \end{center}
\end{table}

\section{Structure of the flow} 
\begin{figure}
  \centerline{\includegraphics[width=6cm]{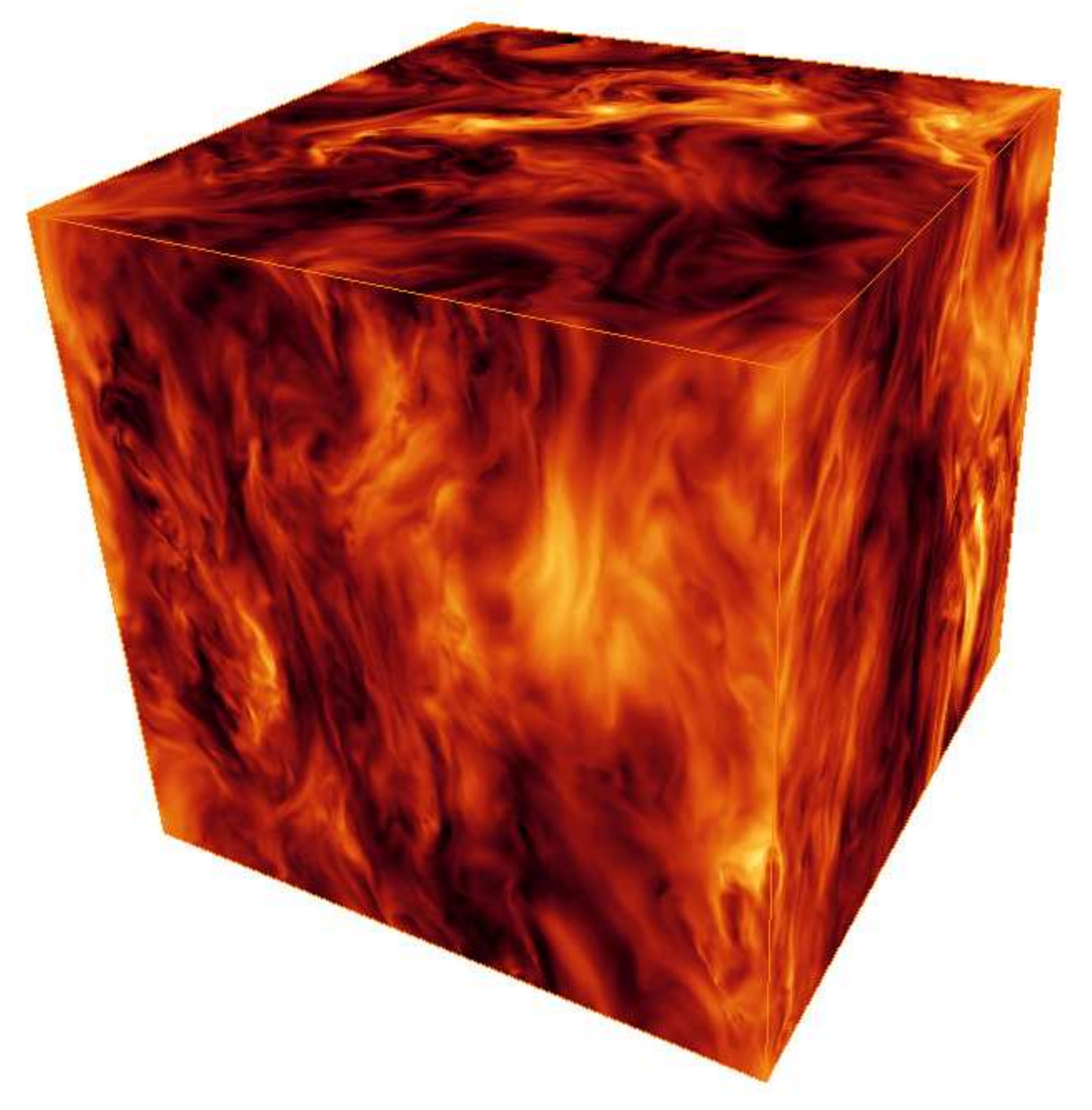} 
              \includegraphics[width=6cm]{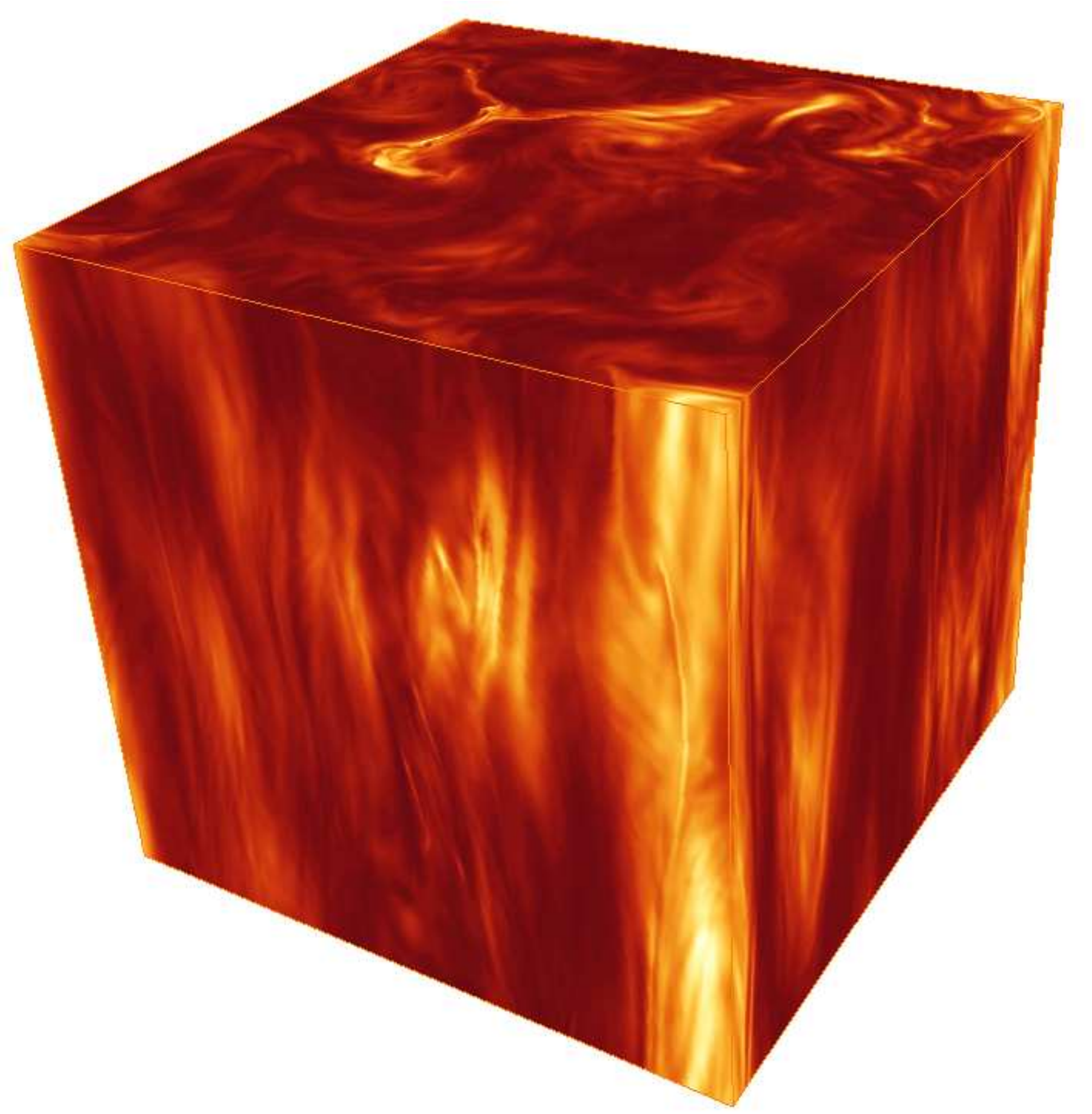}}
    \caption{Visualization of the magnitude of the velocity field $|\boldsymbol{v}(\boldsymbol{r},t)|$ 
             for simulations I (left) and II (right) in the statistically 
             stationary regime of the flow. Brighter colours correspond to higher values of 
             $|\boldsymbol{v}|$ and darker colours to lower values. The rotation axis points in the 
             vertical direction. The pictures show the flow field of the whole $512^3$ simulation domain.}
    \label{fig:flow_visual}
\end{figure}

An immediate impression of the structure of rotating turbulence is
provided by visualizations of the flow field given in figure
\ref{fig:flow_visual}. The system shows a distinct structuring along the
rotation axis which is especially pronounced for simulation II and
indicates a strong velocity correlation along this direction.
The observed trend to a quasi-two-dimensional state with faster
rotation is in agreement with previous experimental findings (see
above).  It is the consequence of a nonlinear analogue of the
Taylor-Proudman theorem \cite[see][]{mahalov_zhou:dimanal,chen_chen_eyink_holm:rotdns128}.
{For more mathematically oriented works on, e.g.,  the dynamics of 
two- ($k_\parallel=0$) and three-dimensional ($k_\parallel\neq 0$) 
flow components as well as the issues of their non-vanishing mutual interaction in the limit
$\Omega\rightarrow\infty$ see 
\cite{babin_mahalov_nicolaenko:regularity} and references therein.} 

The clearest signature of the turbulent energy cascade is found in  
energy spectra and energy flux functions.
Due to the above-mentioned anisotropy of the turbulence, one dimensional spectral
quantities are considered in directions parallel and perpendicular to the rotation axis. 
The axis-parallel nonlinear energy
flux over wavenumber $k_\parallel$ is given by  
$T_{k_\parallel}=\int_0^{k_\parallel} \mathrm{d}k_\parallel'
\int \mathrm{d}k'\int \mathrm{d} k_\perp \left( i \boldsymbol{\omega}^* \cdot
(\boldsymbol{k} \times \widetilde{ \left[ \boldsymbol{v} \times
\boldsymbol{\omega} \right]}_{\boldsymbol{ \scriptscriptstyle k}})
+c.c.\right)/\boldsymbol{k}^2$, with $\widetilde{[\bullet]}$ denoting
Fourier transformation, $\boldsymbol{k}_\perp\perp\boldsymbol{\Omega}$,
$\boldsymbol{k_\parallel'}\parallel \boldsymbol{\Omega}$, and $\boldsymbol{k'} \perp
\boldsymbol{k_\perp},\boldsymbol{k_\parallel'}$. The axis-perpendicular flux function $T_{k_\perp}$
is defined analogously using the integration $\int_0^{k_\perp}\mathrm{d}k'_\perp\int\mathrm{d}k'
\int\mathrm{d}k_\parallel$ with the $k_\perp$-component running along an arbitrary fixed direction in the axis-perpendicular plane. 
The fluxes are normalized with the total
energy dissipation rate $\varepsilon=\mu\int\mathrm{d}V \omega^2$.

{The parallel and perpendicular energy fluxes for $\Omega=0,5,50$  
are negative at all wavenumbers indicating direct energy transfer towards
small scales \cite[][]{mueller_thiele:rotscal}.} 
{The two-dimensionalization visible in figure \ref{fig:flow_visual} 
is not in contradiction with the observed direct fluxes of energy which are a mere consequence 
of the large scale driving of the flow.}
The fluxes weaken with increasing $\Omega$, a trend that is much stronger in
$T_{k_\parallel}$ than in $T_{k_\perp}$.
The increasing anisotropy of
the energy cascade with growing $\Omega$ reflects the nonlinear nature
of the dynamical process that is at least partially responsible for the observed
two-dimensionalization of the flow. The classical picture of energy
transfer through eddy breakup relates a stronger eddy scrambling
perpendicular to the rotation axis than in the parallel direction
along which coherent structures remain comparatively intact.

\begin{figure}
  \centerline{\includegraphics[width=7cm]{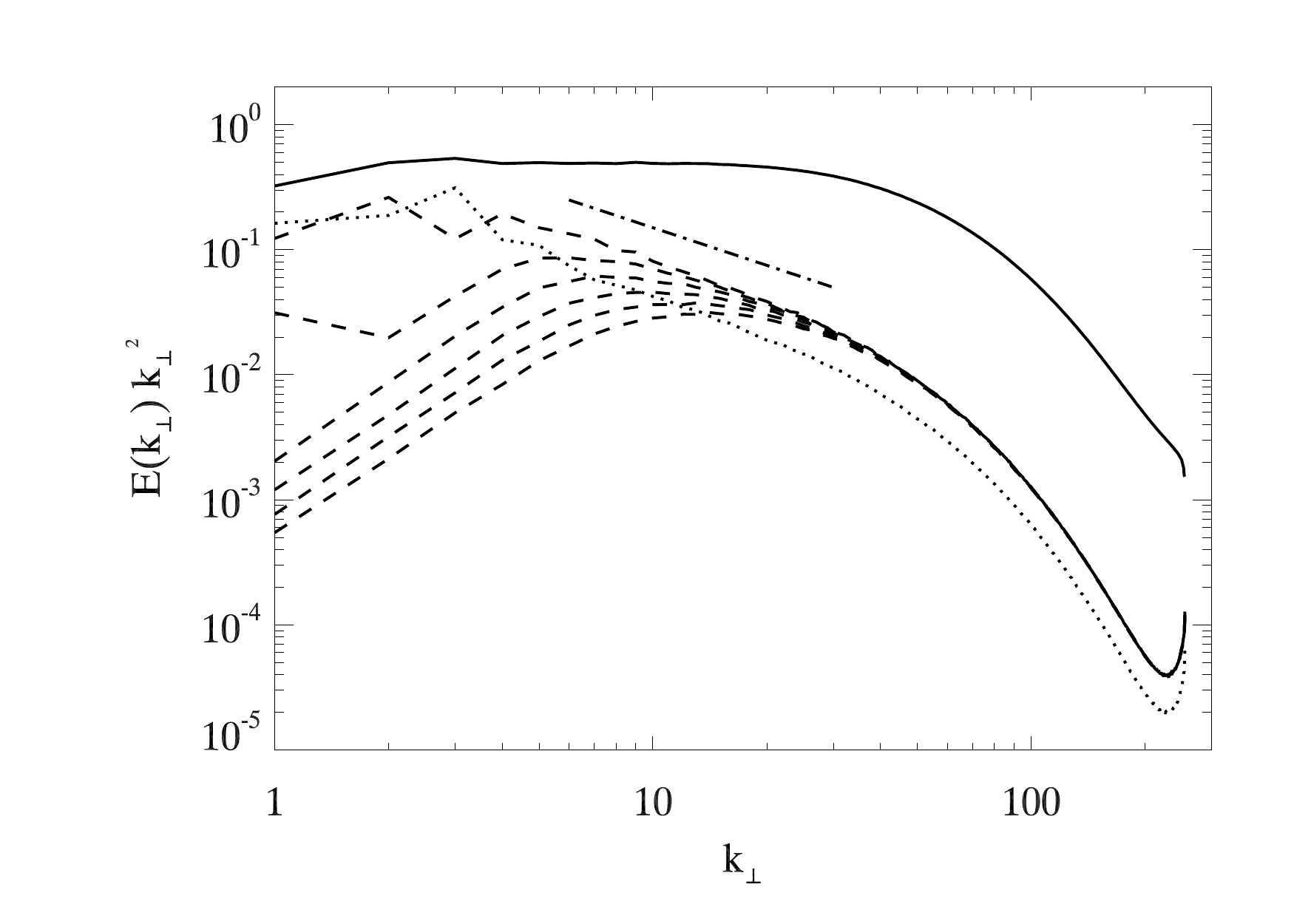}\includegraphics[width=7cm]{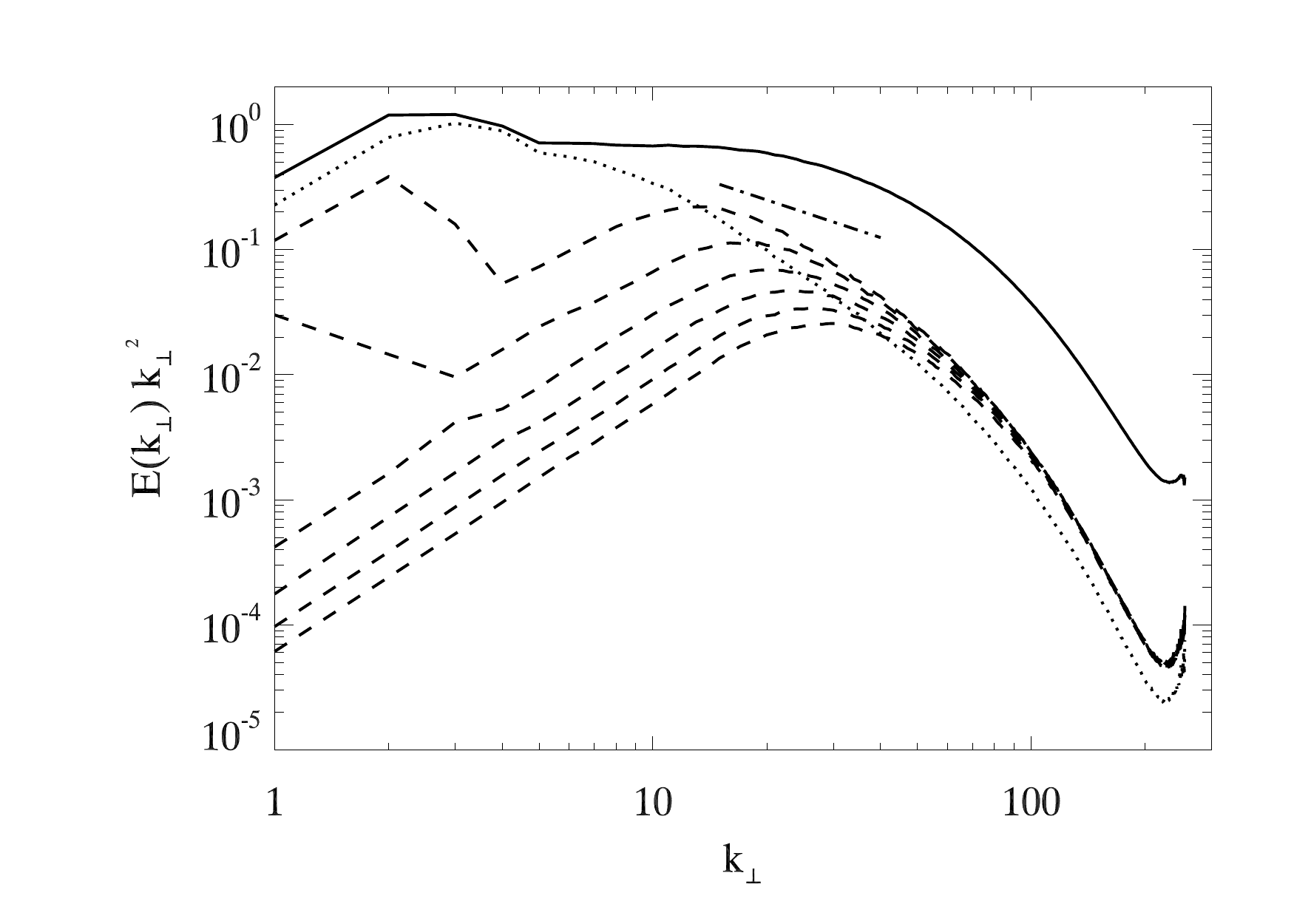}}
    \caption{Perpendicular two-dimensional spectra $E(k_\parallel,k_{\bot})$ for $\Omega = 5$ (left) and $\Omega = 50$ (right) at fixed $k_\parallel$ and compensated with $k_{\bot}^2$. The spectrum for $k_\parallel = 0$ (dotted) differs markedly from those for $k_\parallel = 1-6$ (dashed lines from top down) and higher. Three-dimensional perpendicular spectra {are} obtained by $k_\parallel$-integration from $E(k_\parallel,k_{\bot})$ (solid). The dash-dotted line indicates $k_\perp^{-3}$-scaling.}
    \label{fig:2d_spec}
\end{figure}

The perpendicular energy spectra $E(k_{\bot}) = \int
dk_{\parallel}' \int dk' \left| v_{\boldsymbol{k}}\right|^2/2$ 
shown as solid lines in figure \ref{fig:2d_spec} display inertial range scaling for $4\lesssim
k_\perp \lesssim 20$ (I) and $5\lesssim k_\perp \lesssim 16$ (II). The
shortening of the scaling range with increasing rotation rate is due
to an extension of the dissipation region towards smaller $k$ (see
below). In simulation II a bump persists around $k_f$, where the
forcing region descends into the freely evolving range of scales. It
is a result of the simple forcing scheme and is tolerable at the
chosen numerical resolution as it is not significantly perturbing the
flow beyond $k\approx5$.

In spite of the shortened inertial range in simulation II both cases
clearly exhibit the scaling $E(k_{\bot}) \sim k_{\bot}^{-2}$. These
results agree with the direct numerical simulations of
\cite{yeung_zhou:rotdns256} and \cite{smith_waleffe:rotdns200}(for
$k>k_f$) as well as  shell-model calculations
\cite[see][]{hattori_rubinstein_ishizawa:rotshell}. Recent
experimental results on decaying turbulence subject to rotation
\cite[see][]{morize_moisy_rabaud:exp} yield an energy scaling
exponent $\approx -2.5$ for the micro Rossby number of
$\mathsf{Ro}_\omega=\langle\omega_3^2\rangle^{1/2}/(2\Omega)\simeq0.08$
(corresponding to our case II) and an exponent of $\approx -1.7$ for
$\mathsf{Ro}_\omega\simeq 0.7$ (corresponding to I). The experimental
configuration with an initial excitation of the flow followed by decay
under rotation is however not directly comparable to the simulations
described here, which are characterized by a continuous large-scale
forcing. {We note in passing that these Rossby numbers place the present simulations
in the ``intermediate'' Rossby number regime of \cite{bourouiba_bartello:rotdecompo}.}
In the experiment of \cite{baroud_etal:expspec} the same
scaling $\sim k^{-2}$ is observed although here forcing at the small
scales produces an inverse energy cascade.

Several theoretical models arrive at the scaling shown in
Fig. \ref{fig:2d_spec}: In the context of quasi-normal closure theories
\cite{zhou:dimanal} and \cite{mahalov_zhou:dimanal} have conducted a
dimensional analysis of the energy flux terms, assuming $\tau^{\ast}
\sim \tau_{\Omega} \sim \Omega^{-1}$ for the relaxation timescale of
nonlinear interactions $\tau^{\ast}$, whereas
\cite{canuto_dubovikov:formal} have analyzed the energy flux
formulated in helical mode decomposition, both arriving at the same
scaling result. On the contrary, weak turbulence theory yields $E(k)
\sim k^{-3}$ \cite[see][]{cambon_rubinstein_godeferd:rotweak,bellet_etal:weaksimul} in the 
asymptotic quasi-normal Markovian approximation valid for $k_\parallel \neq 0$ and
$E(k) \sim k^{-2}$ \cite[see][]{galtier:rotweak}. The former result
is regarded as the consequence of a strongly anisotropic spectral energy distribution
with respect to $\boldsymbol{\Omega}$
while the latter explicitly assumes spectral isotropy in combination with a phenomenological 
nonlinear transfer time based on the wave-kinetic equation. Generally, however, wave turbulence 
theory predicts anisotropic scalings in the context of rotating turbulence.

\begin{figure}
  \centerline{\includegraphics[width=10cm]{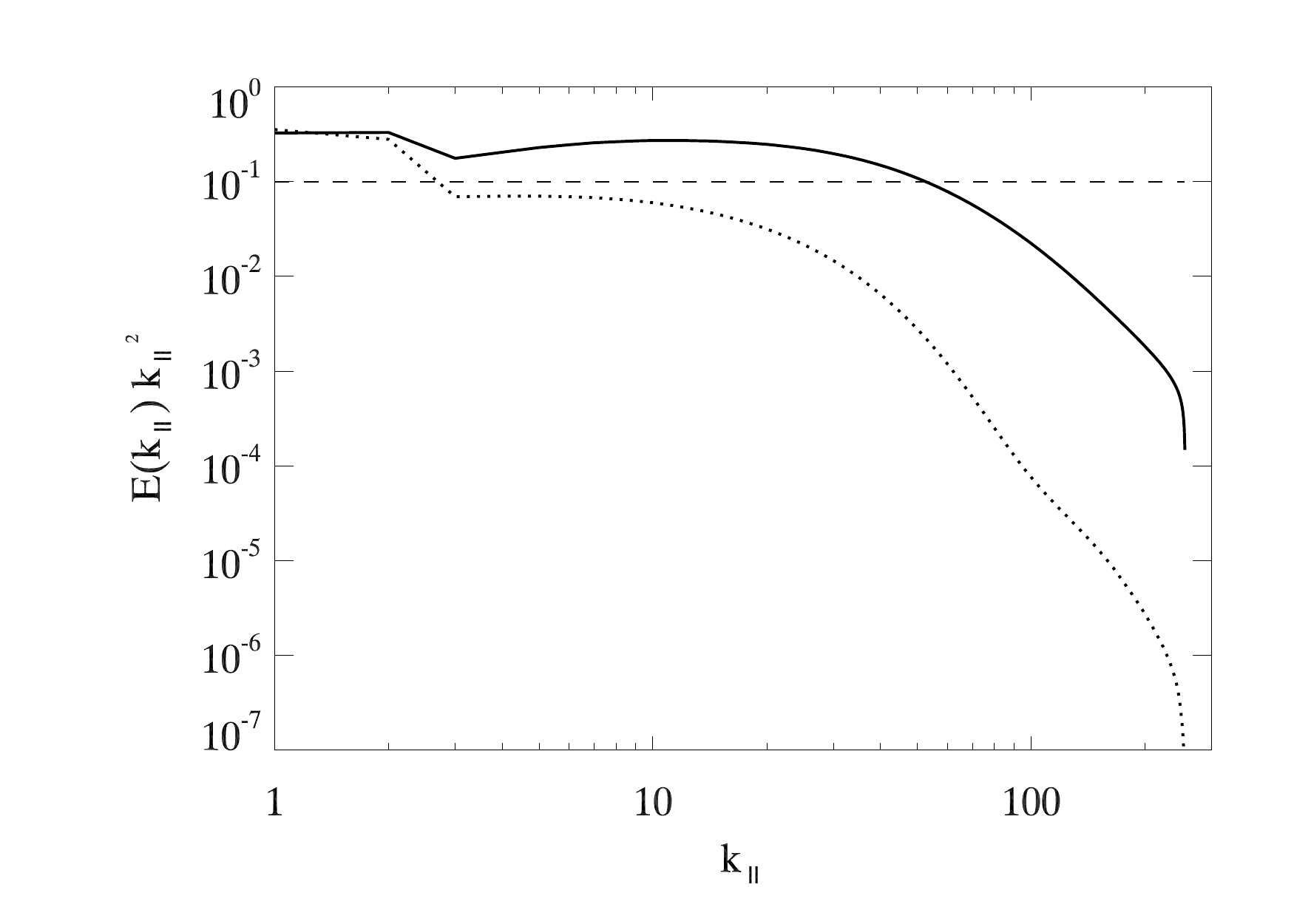}}
    \caption{One-dimensional energy spectra parallel to the rotation axis compensated with $k_{\parallel}^2$ for $\Omega = 5$ (solid), and $\Omega = 50$ (dotted).}
    \label{fig:par_spec}
\end{figure}
It is noteworthy that the axis-parallel spectra shown in figure \ref{fig:par_spec} do not show clear
scaling. This indicates that the perpendicular domain of
wavenumbers provides the dominant contribution to the angle-averaged
scaling $E(k) \sim k^{-2}$ found in some of the above-mentioned
numerical and experimental studies.  

The contributions to the 
perpendicular spectrum for fixed $k_\parallel$, shown in figure
\ref{fig:2d_spec}, exhibit for $\Omega=5$ $k^{-3}_\perp$-scaling in the vicinity of $k_\parallel=0$. 
Note that 
$E(k_\perp)\sim \int dk_\parallel E(k_\parallel,k_\perp)\neq E(k_\parallel=const,k_\perp)$.
The scaling range, however, shrinks from the 
low $k$ side with growing $k_\parallel$. For $\Omega =
50$ the spectral contributions for fixed $k_\parallel$ show no clear scaling. Evidently, the 
observed $k^{-2}_\perp$ power-law is due to spectral contributions from all wavenumbers
$k_\parallel$. This observation is in contradiction
to weak turbulence simulations reported in \cite{bellet_etal:weaksimul} which give $k^{-2}_\perp$-scaling in directions perpendicular to $\boldsymbol{\Omega}$ and suggest that $k^{-3}$-spectra are generated by spherical averaging. 
Furthermore and especially for $\Omega =
50$, the energy is mainly concentrated around the plane
$k_\parallel=0$ as expected from previous studies
\cite[see, e.g.,][]{cambon_lacquin:edqnm3,cambon_mansour_godeferd:etransfer}
and in agreement with the observed nonlinear two-dimensionalization. 

{ 
The observed axis-perpendicular spectral scaling 
can be described by a simple
phenomenology of rotating turbulence proposed in
\cite{mueller_thiele:rotscal} which tries to capture the effect of
inertial oscillations on convective fluid motion.  The phenomenology
applies in the inertial-range under the additional constraint
$k_\Omega>k>\min(k_\mathrm{d},k_\mathrm{d}^\Omega)$, i.e. when
nonlinear energy transfer is dominated by Coriolis force effects.
This spectral interval is bounded by the rotation wavenumber 
$k_\Omega=(\Omega^3/\varepsilon)^{1/2}$ whose inverse, also called {Rossby deformation radius}, 
is analogous to the geophysical Ozmidov scale \cite[see e.g.][]{ozmidov:length}, the classical
Kolmogorov dissipation wavenumber $k_d=(\varepsilon/\mu^3)^{1/4}$, and its
rotation-modified counterpart
$k_d^\Omega=(\varepsilon/\Omega)^{1/2}/\mu$ \cite[see, e.g., 
also][]{zeman:rotscale, canuto_dubovikov:formal}.}

{
For turbulent fluctuations $v_\xi$ perpendicular to
$\boldsymbol{\Omega}$ on spatial scale $\xi\sim k_\perp^{-1}$ the
influence of rotation on the inertial-range energy transfer in planes
perpendicular to $\boldsymbol{\Omega}$ is phenomenologically described
similarly to the Iroshnikov-Kraichnan picture of magnetohydrodynamic
turbulence \cite[][]{iroshnikov:ikmodel,kraichnan:ikmodel}.  Taking
into account the random displacements that a fluid particle experiences 
during a nonlinear interaction of turbulent
eddies under the influence of inertial oscillations leads to a prolonged energy transfer time $\tau_\mathrm{tr}\sim
\tau_{\mathrm{NL}_\perp}^2/\tau_\Omega$ with
$\tau_{\mathrm{NL}_\perp}\sim\xi/v_\xi$ as compared to the non-rotating case. This yields by standard dimensional
arguments the non-intermittent scaling relation
$v_\xi\sim(\Omega\varepsilon)^{1/4}\xi^{1/2}$ corresponding to the
observed scaling of the energy spectrum $\sim k_\perp^{-2}$ \cite[see
also][]{zhou:dimanal,canuto_dubovikov:formal}.
}

{
The numerical values of the characteristic 
wave numbers are for simulation (I): $k_\Omega\simeq 50$, $k_d^\Omega\simeq 250$, $k_d\simeq
167$, and for simulation (II): $k_\Omega\simeq 1581$, $k_d^\Omega\simeq 79$, $k_d\simeq
167$.} 
It should be noted that, according to their definition, for
increasing $\Omega$, $k_\Omega$ grows while $k_d^\Omega$
decreases. This is the reason for the smaller inertial range in system I and
could also explain why in earlier works with lower resolution the
scaling exponent of the energy spectrum for high rotation rates has
been difficult to pin down.

The two-dimensionalization evident in figure
\ref{fig:flow_visual}, due to the large-scale driving, does not lead
to an inverse energy cascade but still leaves some spectral traces
that can be understood by the following ordering considerations: it
is proposed that for fixed $\Omega$ the are three
different regimes of rotating turbulence: a) turbulence
with negligible rotation effects at $k \gg k_\Omega$, 
b) turbulence strongly dominated by the Coriolis force
at $k \ll k_\Omega$ which is asymptotically described by
weak-turbulence theory with the associated purely nonlinear
two-dimensionalization via anisotropic nonlinear energy transfer, and
c) rotating turbulence in the transient regime at $k\lesssim k_\Omega$
observed in the present simulations where
quasi-two-dimensionalization occurs due to the ``nonlinear Taylor-Proudman
theorem'' (see above). 
{
This effect entails an observable energetic separation of 2D ($k_\parallel=0$) and 3D
($k_\parallel \neq 0$) components of the velocity field which exhibit 
an energetically weaker coupling at high rotation rate \cite[see,
e.g.,][and references therein]{bourouiba_bartello:rotdecompo}. It can, however, be shown 
rigorously \cite[see e.g.][]{babin_mahalov_nicolaenko:regularity} that the dynamical coupling of 
2D and 3D fluctuations remains finite for all values of $\Omega$.}  In
regime a) turbulence is based on nonlinear eddy interaction, regime b)
represents turbulence governed by (weak) wave interaction, while
nonlinear dynamics underlying regime c) correspond to wave-modified
eddy interaction. In regime c) the 2D fluctuations should exhibit
dynamics different from the 3D fluctuations as indicated by
the nonlinear Taylor-Proudman theorem. {A possible scenario that would be 
in agreement with the observed scaling laws consists of  a direct cascade of  3D energy
as described by the proposed
phenomenology in combination with an inverse cascade of 2D energy 
(towards wavenumbers smaller than the forcing range) or a
direct enstrophy cascade (towards wavenumbers larger than the forcing
range). In this picture the $k^{-3}$ inertial-range scaling of 2D energy
could be interpreted as the signature of a direct 2D enstrophy cascade.}

{
Regime c), in fact, describes a state of continuous transition from $\mathsf{Ro}\rightarrow\infty$ (negligible rotation) 
to $\mathsf{Ro}\rightarrow 0$ (weak inertial-wave-turbulence).
This transitional state is applying to most realistic rotating systems. The dynamical phenomenology 
by \cite{mueller_thiele:rotscal} could be a reasonable starting point for extending the theoretical description
of rotating turbulence to include the growing influence of inertial waves on turbulence dynamics.} 
 
To obtain the scaling exponents $\zeta_p$ of the axis-perpendicular
longitudinal velocity structure functions,
$S_p=\langle|[\boldsymbol{v}(\boldsymbol{r})-\boldsymbol{v}(\boldsymbol{r}+\boldsymbol{\xi})]\cdot\boldsymbol{\xi}/\xi|^p
\rangle\sim\xi^{\zeta_p}$ their extended self-similarity (ESS) first
observed by \cite{benzi:ess} in the non-rotating case is exploited. By
this heuristic but generally accepted procedure exponents up to order
$p=8$ can be determined with sufficient precision in spite of the
limited spatial resolution and overall duration of the simulations.
The velocity field is resolved into a 2D component ($k_\parallel=0$)
and 3D fluctuations ($k_\parallel\neq 0$) \cite[see,
e.g.,][]{bourouiba_bartello:rotdecompo}.  The relative exponents
$\zeta_p/\zeta_2$ obtained via ESS from the 3D velocity coincide with
the $\zeta_p$ since the numerical results $E_{k_\perp}\sim
k_\perp^{-2}\sim k_\perp^{-(\zeta_2+1)}$ yield $\zeta_2^\mathrm{3D} =
1$. {For the 2D component $\zeta_2^\mathrm{2D}=2$ is chosen consistently with
the inertial-range scaling of the 2D energy spectrum
$\sim k_\perp^{-3}$ (dotted line in figure \ref{fig:2d_spec}).}

\begin{figure}
  \centerline{\includegraphics[width=7cm]{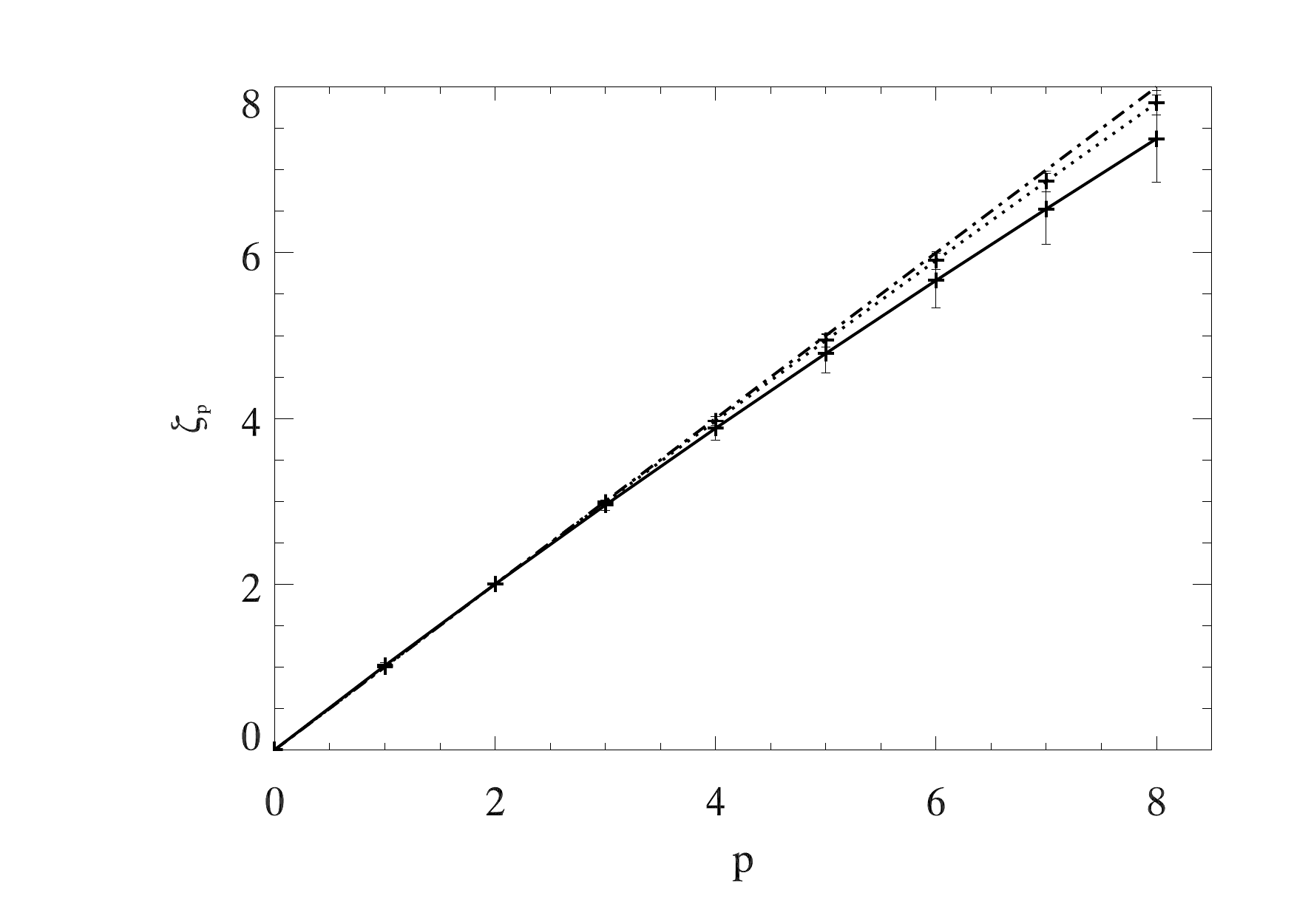}\includegraphics[width=7cm]{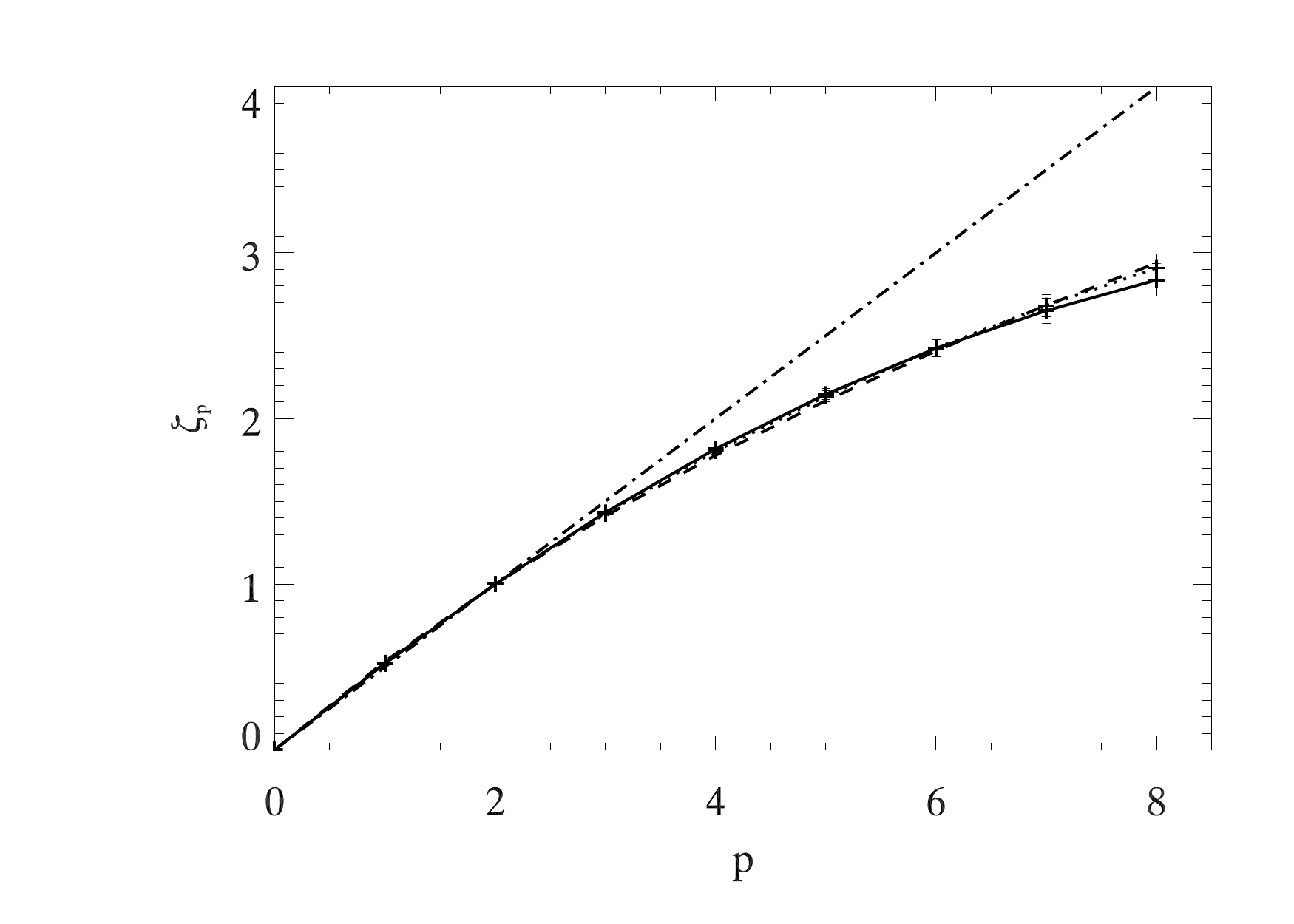}}
    \caption{Axis-perpendicular structure function scaling exponents $\zeta_p$ of velocity fluctuations with $k_\parallel=0$ (left) and $k_\parallel \neq 0$ (right) up
to order 8 for moderate, strong, and no rotation. Solid: $\Omega=5$, dotted: $\Omega=50$, dash-3-dot: $\Omega=0$, dashed:
intermittency model (\ref{intermodel}), dash-dotted: non-intermittent scaling, $\zeta_p=p/2$
($\Omega=5$) and $\zeta_p=p$ ($\Omega=50$).}
    \label{fig:strfun}
\end{figure}

The ESS-results shown in figure \ref{fig:strfun} and table \ref{tab:strucs} indicate a
{weakly} intermittent structure of the 2D velocity
component.{ The difference between exponents from non-rotating turbulence and those for $\Omega=50$ indicates that the  
reduction of intermittency for higher $\Omega$ is not a simple consequence of the removal of fluctuations with 
higher $k_\parallel$.}  The 3D fluctuations 
exhibit an intermittent signature that is up to order $p=7$ virtually
independent of the rotation rate. The highest order scaling 
exponent is, however, suggesting slightly higher intermittency for $\Omega=5$ than
for $\Omega=50$.

{
A comparison with two-point statistics obtained with the total turbulent velocity field including all $k_\parallel$ contributions
\cite[][]{mueller_thiele:rotscal} reveals that the 2D-component of velocity tends to govern 
structure function scaling for $\Omega=50$. In contrast, the $\zeta_p$ taken with $\Omega=5$ are rather close to the 
values obtained with the $k_\parallel\neq 0$ velocity. The dominance of the 2D fluctuations in 
structure function scaling at high $\Omega$ is plausible because of the reduced activity 
of axis-perpendicular nonlinear interactions that is visible, e.g., in the corresponding energy flux \cite[][]{mueller_thiele:rotscal}.   
The accompanying reduction of intermittency in the 2D flow component with higher rotation rate points towards
stronger dynamical influence of inertial waves since weak wave-turbulence is generally non-intermittent. This effect
is stronger in the 2D velocity than in the 3D component as the nonlinear energy flux is strongly depleted in
the axis-parallel direction at high rotation rate compared to the axis-perpendicular flux (\textit{ibid.}).   
} 
 
{The 3D structure function scaling is well represented by an intermittency model
based on the She-L\'ev\^eque (SL) ansatz
\cite[][]{she_leveque:model}. The SL-model is quite successful in
describing intermittent scalings of two-point statistics in various
non-rotating turbulent flows. It can be written in a general form
exhibiting two parameters which can be determined by physical
considerations \cite[][]{grauer_krug:mhdsl,pol_pouq:mhdsl,mueller_biskamp:3dmhdscale}:
$\zeta_p=p/(3g)+C_0(1-(1-2/(3C_0))^{p/g})$.  While $C_0$ stands for
the co-dimension of the most singular dissipative structures, $g$ is
connected to the basic nonintermittent scaling exponent, $v_\ell\sim
\ell^{1/g}$. In the present three-dimensional simulations the
most-singular structures are quasi-one-dimensional vortex filaments,
i.e. $C_0=2$, while relation (\ref{rotscal}) yields g=2. This results
in the intermittency model 
\begin{equation}\zeta_p=p/6 + 2(1-(2/3)^{p/2}) \label{intermodel}\end{equation} which
agrees well with the numerical data as shown in figure \ref{fig:strfun} (see also table 
\ref{tab:strucs}).} 
\begin{table}
  \begin{center}
  \begin{tabular}{lllllll}
      $p$   & $\zeta_p^\mathrm{2D, \Omega=5}$ & $\zeta_p^\mathrm{2D, \Omega=50}$&$\zeta_p^\mathrm{nonintermitt}$ & $\zeta_p^\mathrm{3D, \Omega=5}$ & $\zeta_p^\mathrm{3D, \Omega=50}$   & $\zeta_p^\mathrm{model}$  \\[3pt]
       1   &$1.02\pm0.03$&$1.0\pm0.01$ & 1 &$0.52\pm0.003$ & $0.52\pm0.002$ & $ 0.53 $  \\
       2   & $2$         & $2$         & 2 &$1$            & $1$            & $ 1 $  \\
       3   &$2.95\pm0.06$&$2.99\pm0.02$& 3 &$1.43\pm0.006$ & $1.43\pm0.007$ & $ 1.41 $  \\
       4   &$3.88\pm0.1$ &$3.97\pm0.05$& 4 &$1.82\pm0.02$  & $1.80\pm0.02$  & $ 1.78 $  \\
       5   &$4.78\pm0.2$ &$4.94\pm0.08$& 5 &$2.15\pm0.03$  & $2.13\pm0.03$  & $ 2.11 $  \\
       6   &$5.67\pm0.3$ &$5.91\pm0.1 $& 6 &$2.42\pm0.05$  & $2.42\pm0.05$  & $ 2.41 $  \\
       7   &$6.53\pm0.4$ &$6.86\pm0.1 $& 7 &$2.65\pm0.07$ & $2.68\pm0.07$   & $ 2.68 $   \\
       8   &$7.38\pm0.5$&$7.81\pm0.2  $& 8 &$2.84\pm0.1$ & $2.91\pm0.09$    & $ 2.94 $    \\
  \end{tabular}
  \caption{Structure function scaling exponents for 2D and 3D components of the velocity field with $\Omega=5$ and $\Omega=50$. The values are obtained by ESS assuming $\zeta_2^\mathrm{3D}=1$ and
$\zeta_2^\mathrm{2D}=2$.} 
  \label{tab:strucs}
  \end{center}
\end{table}

\begin{figure}
  \centerline{\includegraphics[width=7cm]{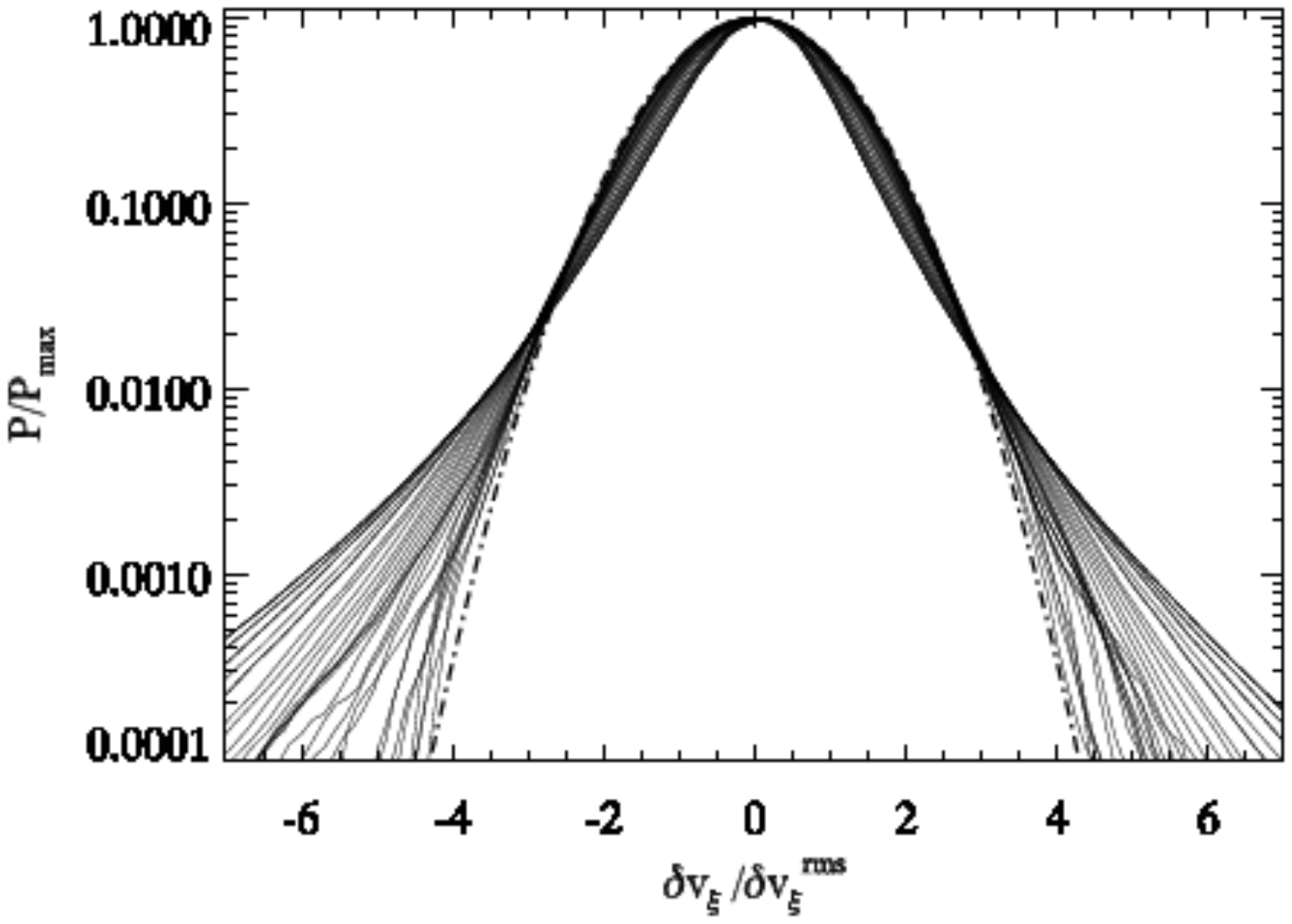}\includegraphics[width=7cm]{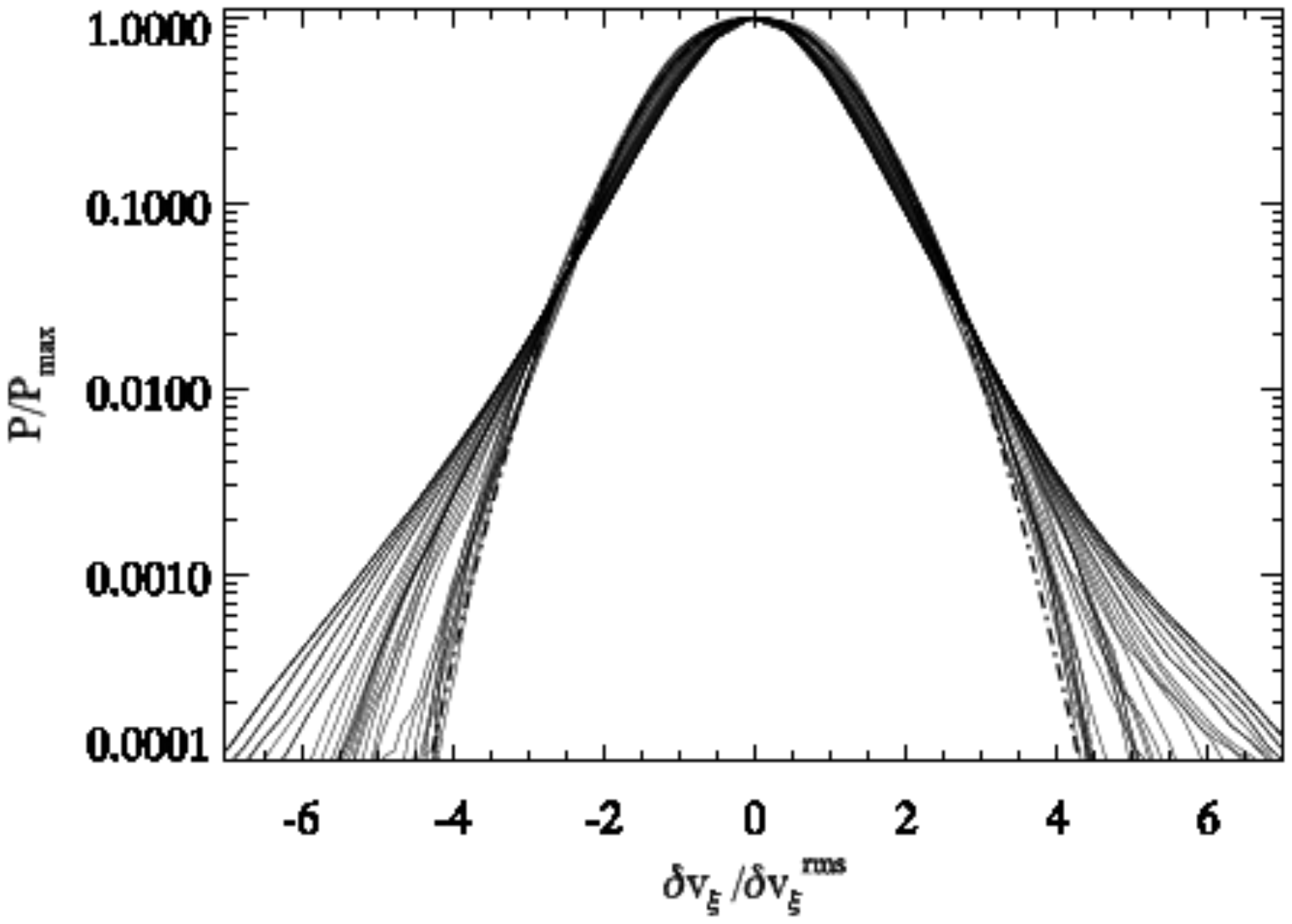}}
    \caption{PDFs of 3D velocity differences ($k_\parallel \ne 0$) taken at different spatial distances
(increasing from left to center in each plot) for $\Omega = 5$ (left) and $\Omega = 50$
(right). The width is rescaled by the velocity variance, and all curves are shifted to a common maximum at 1 to allow for simpler comparison with the shape of a Gaussian (dash-dotted curve).
}
    \label{fig:pdfs}
\end{figure}

\begin{figure}
  \centerline{\includegraphics[width=7cm]{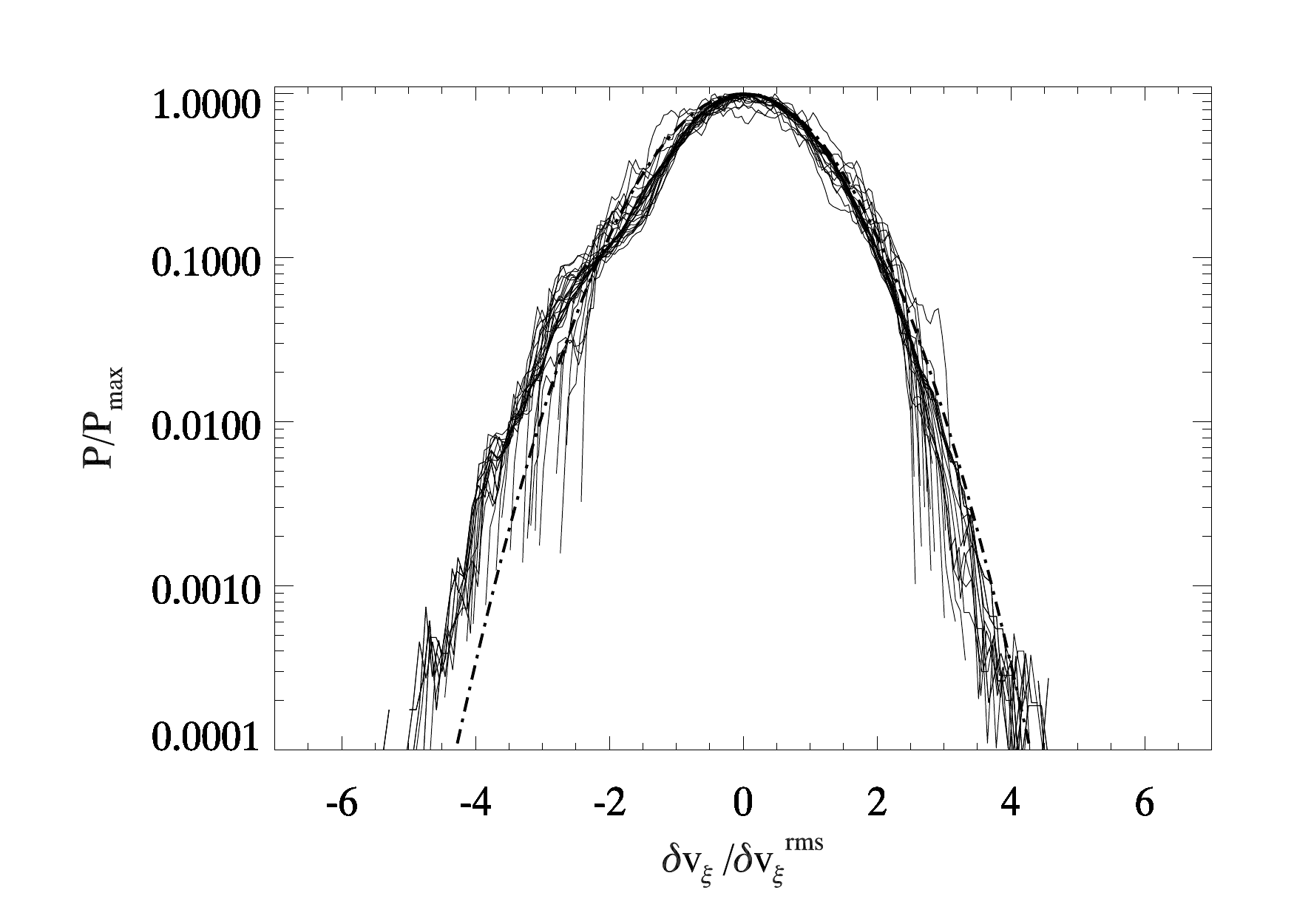}\includegraphics[width=7cm]{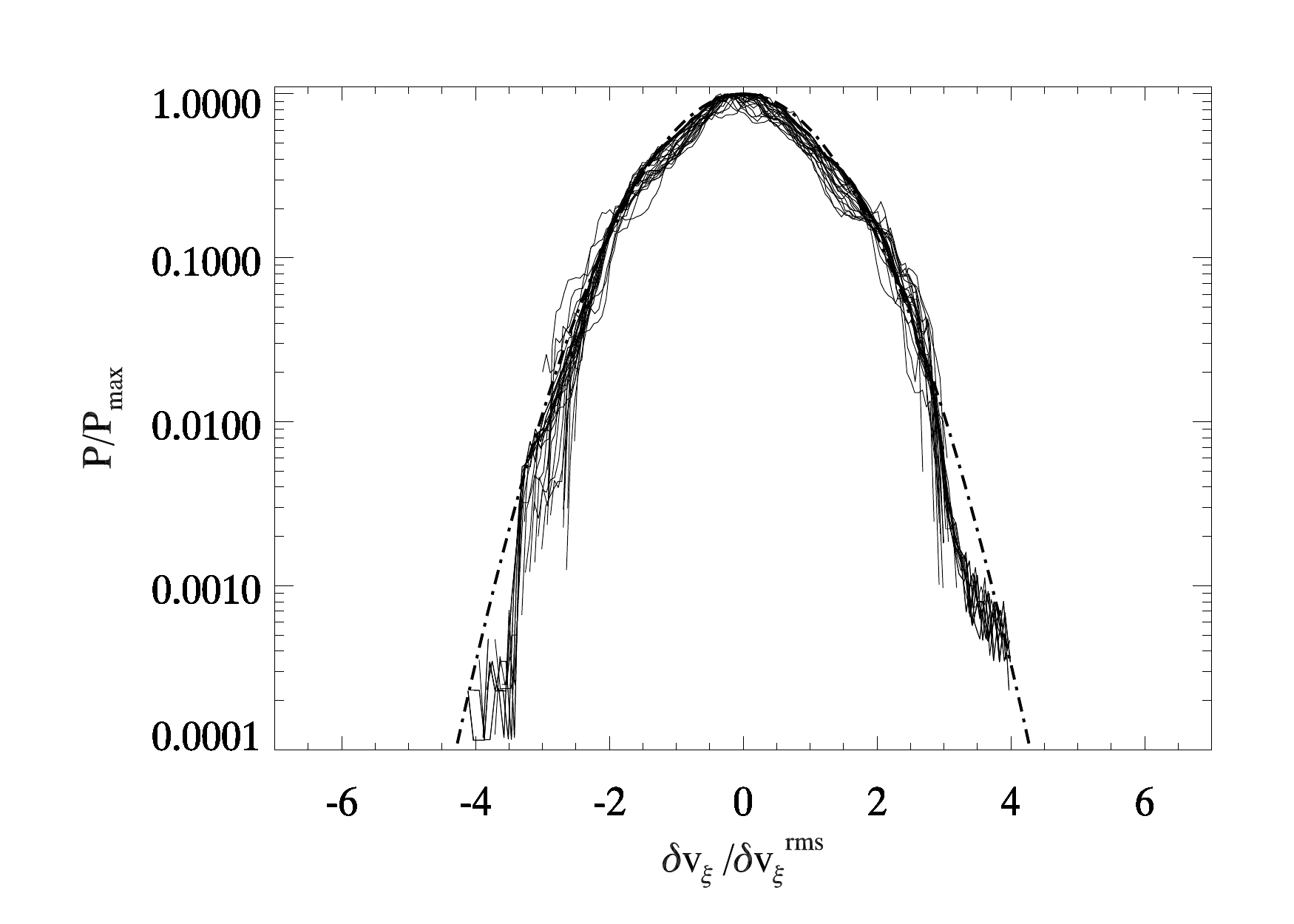}}
    \caption{PDFs of 2D velocity differences ($k_\parallel =0$) taken at different spatial distances for $\Omega = 5$ (left) and $\Omega = 50$
(right). The width is rescaled and all curves are shifted as in figure \ref{fig:pdfs}.  The dash-dotted curve represents a Gaussian.}
    \label{fig:2dpdfs}
\end{figure}

\begin{figure}
  \centerline{\includegraphics[width=7cm]{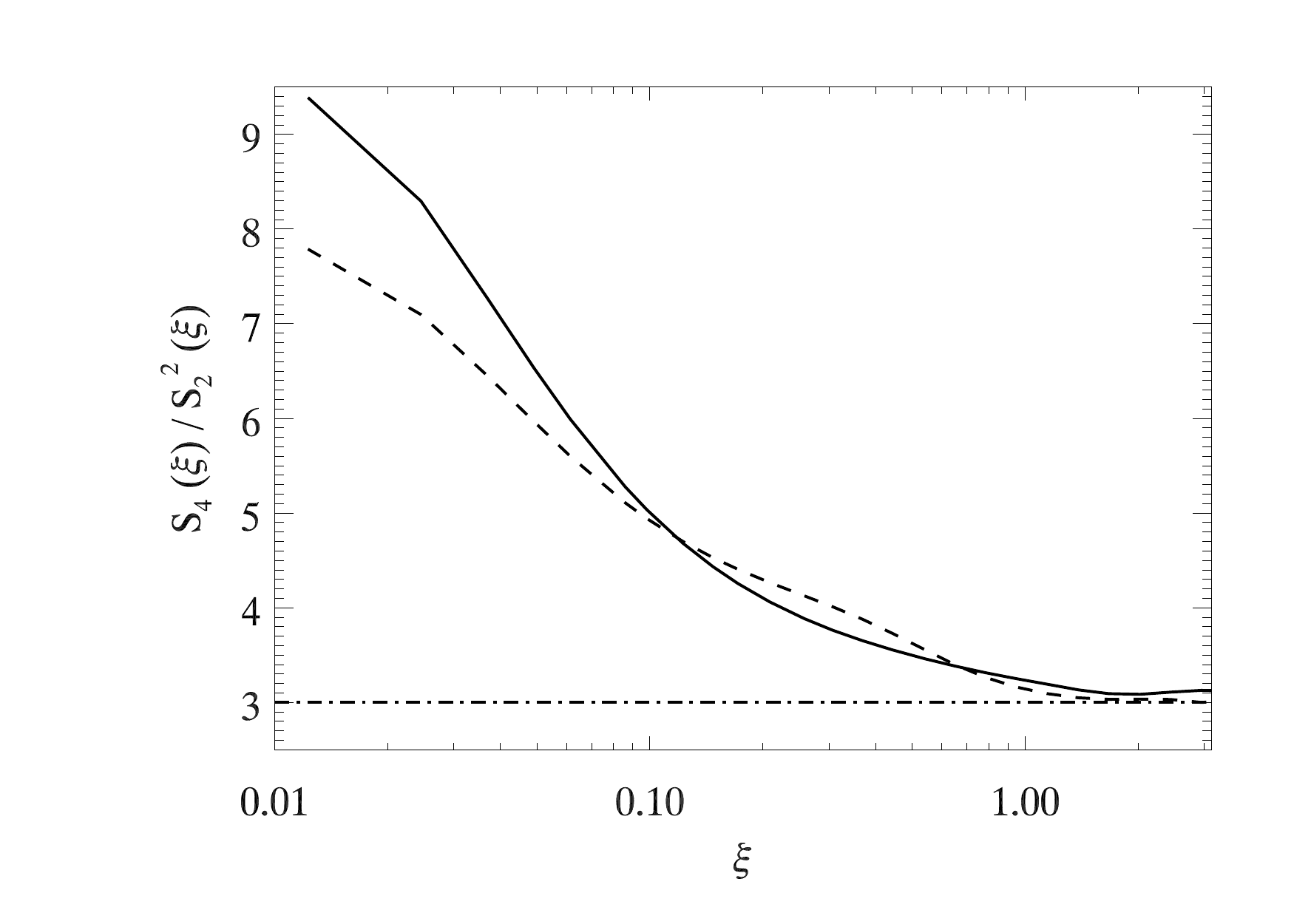}\includegraphics[width=7cm]{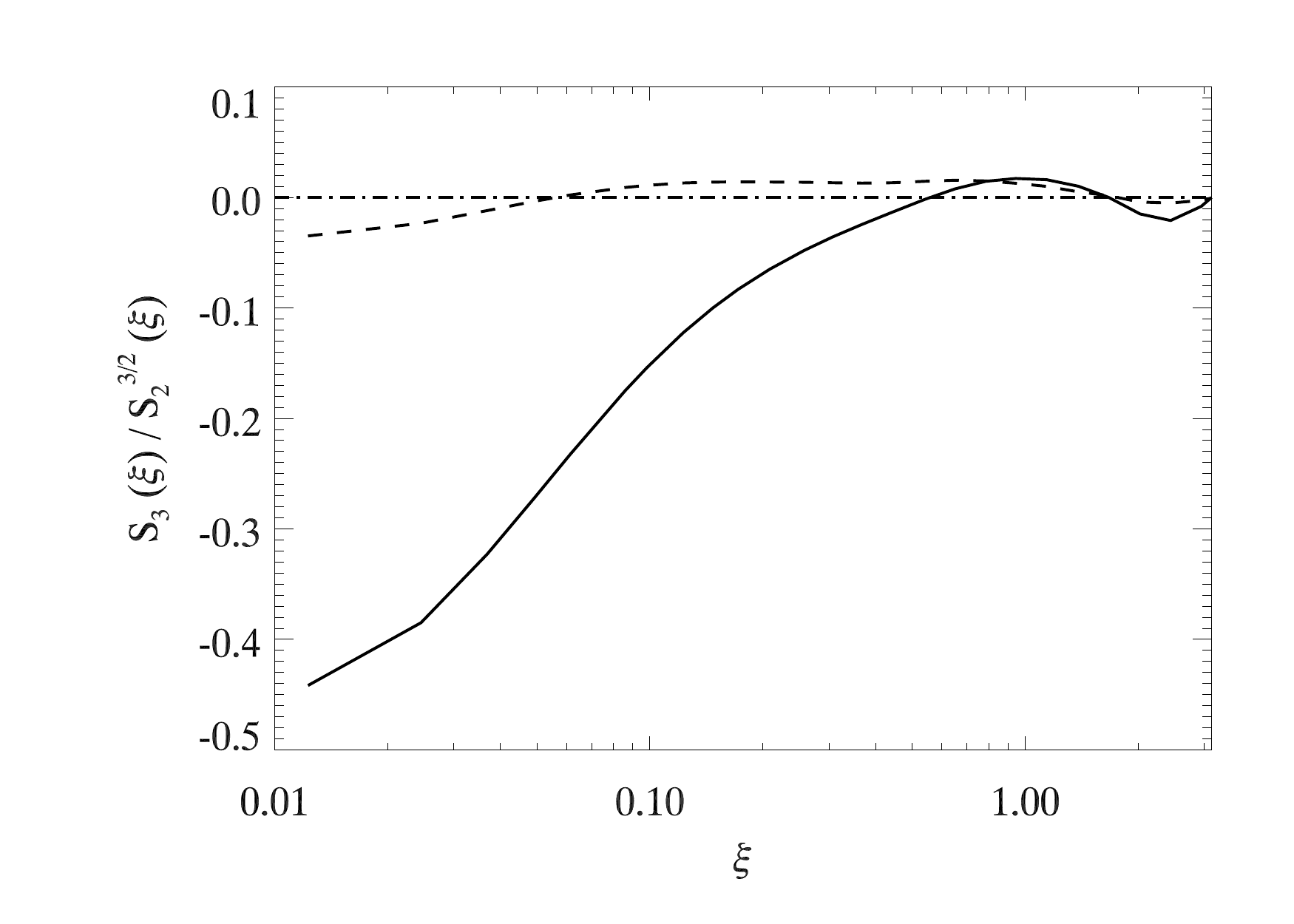}}
    \caption{Time-averaged flatness $S_4(\xi)/(S_2(\xi))^2$ (left) and skewness $S_3(\xi)/(S_2(\xi))^{3/2}$ (right) of the PDFs corresponding to the 3D velocity fluctuations (figure \ref{fig:pdfs}) for for $\Omega = 5$ (solid line), $\Omega = 50$ (dashed line), and $\Omega=0$ (dash-3-dotted line). The dash-dotted lines
display the respective values for a Gaussian.}
    \label{fig:flatness_skewness}
\end{figure}

\begin{figure}
  \centerline{\includegraphics[width=7cm]{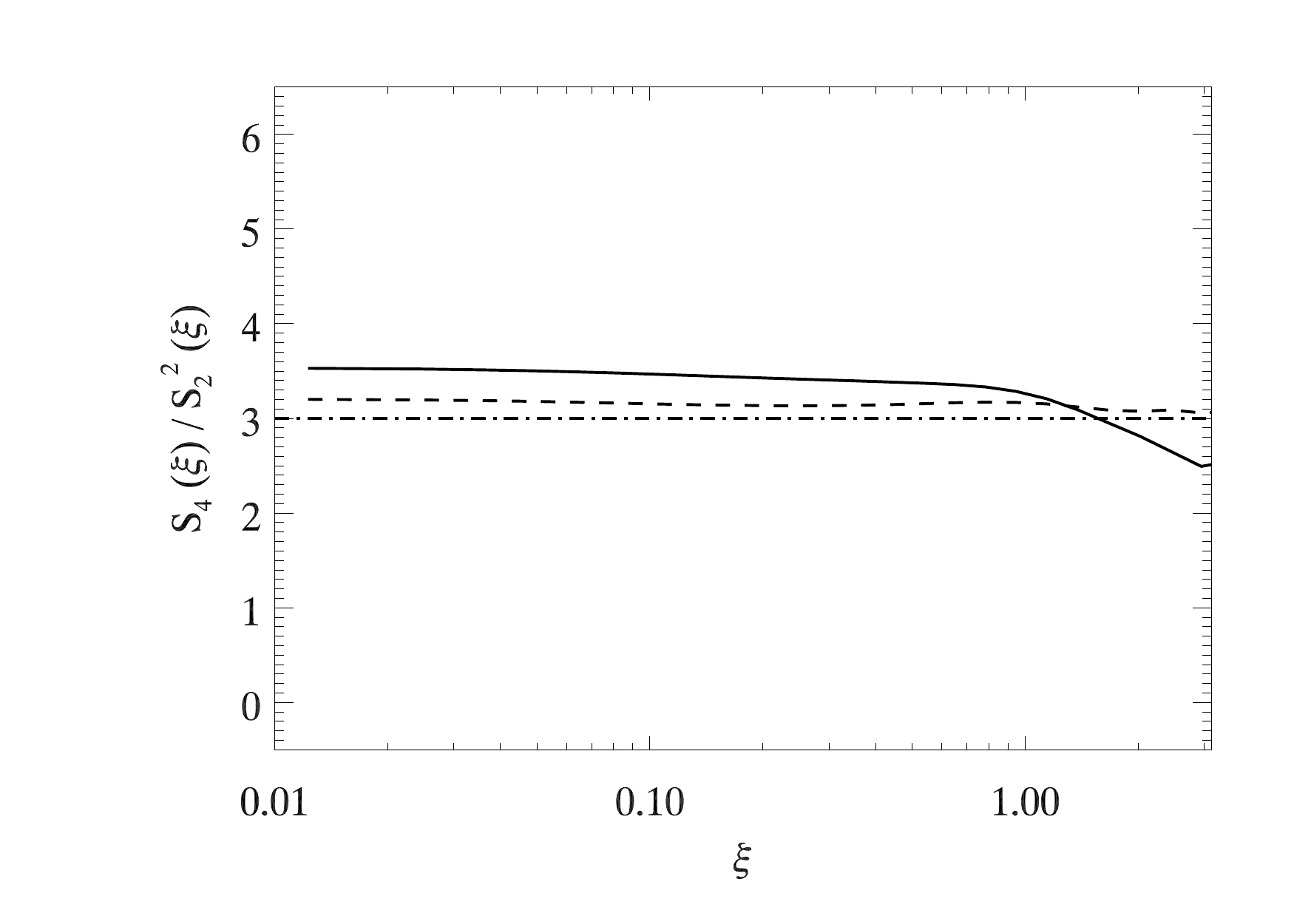}\includegraphics[width=7cm]{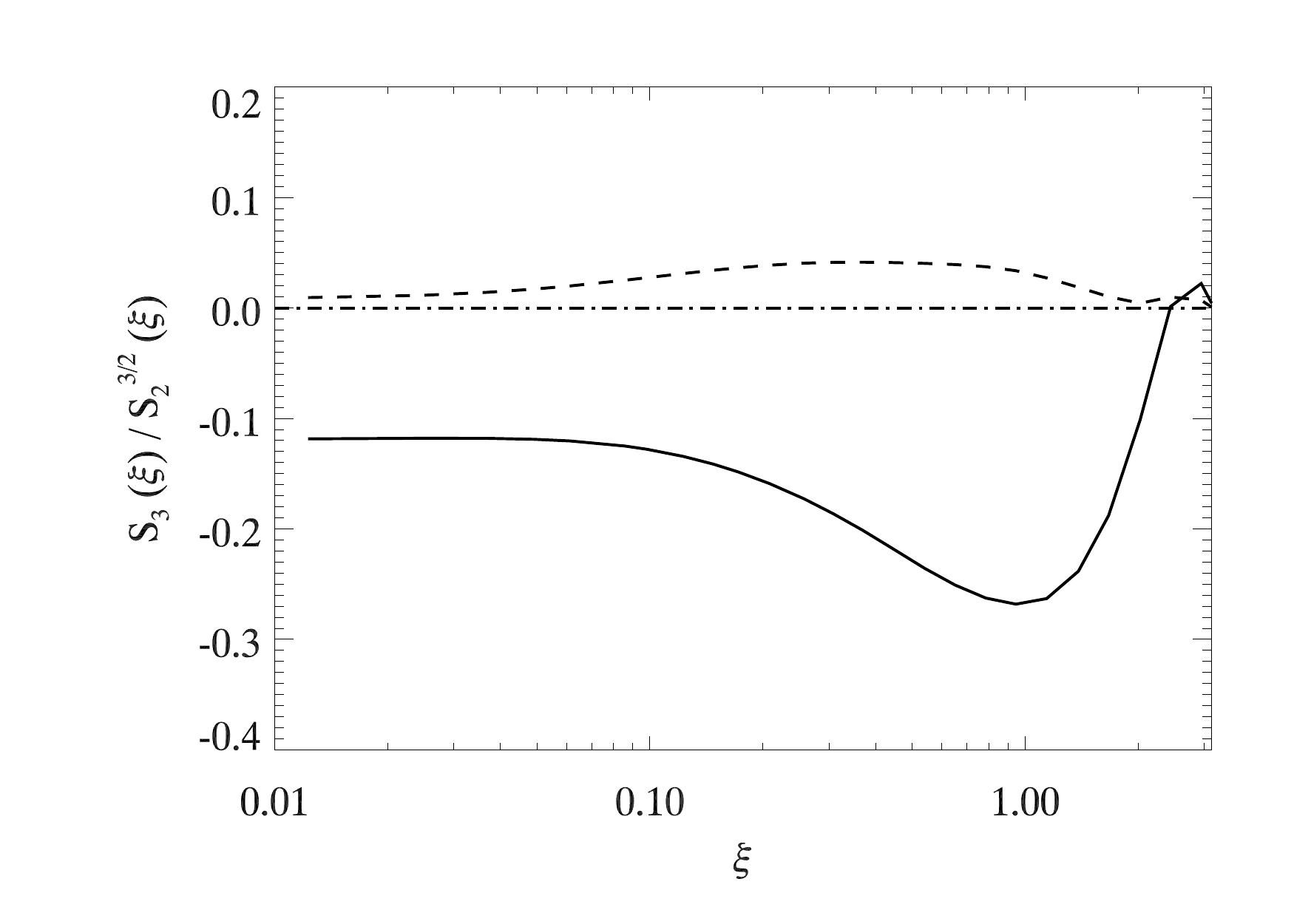}}
    \caption{Time-averaged flatness $S_4(\xi)/(S_2(\xi))^2$ (left) and skewness $S_3(\xi)/(S_2(\xi))^{3/2}$ (right) of the PDFs corresponding to the 2D velocity fluctuations  (figure \ref{fig:2dpdfs}) for for $\Omega = 5$ (solid line), $\Omega = 50$ (dashed line), and $\Omega=0$ (dash-3-dotted line).The dash-dotted lines
display the respective values for a Gaussian.}
    \label{fig:flatskew}
\end{figure}

To conclude the analysis of the structure of rotating turbulence we
present probability {density} functions (PDFs) $P(\delta v_\xi)$ of
the axis-perpendicular longitudinal velocity increments $\delta v_\xi
=
[\boldsymbol{v}(\boldsymbol{r})-\boldsymbol{v}(\boldsymbol{r}+\boldsymbol{\xi})]\cdot\boldsymbol{\xi}/\xi$.
{Characteristic one-time}{ PDFs for the 3D velocity field are shown in figure \ref{fig:pdfs}. For
both rotation rates the PDFs shift from Gaussian form at large
scales to functions with stretched exponential tails at small scales. This well-known
behaviour indicates that both flows are intermittent in
agreement with the structure function scaling. 
The tendency towards nonintermittent quasi-two-dimensional flow for increasing rotation
rate is visible in the slight contraction of the normalized PDFs that
is quantified by decreasing flatness,{ 
$S_4/(S_2)^2$, taking on values of $9.4$ ($\Omega=5$) and of $7.8$ 
($\Omega=50$) at small scales $\xi$. At large scales the flatness is, for both values of 
$\Omega$, close to $3$, a value 
characteristic for a Gaussian distribution. The skewness $S_3/S_2^{3/2}$ is also 
consistent with Gaussian statistics at large scales where it is ranging around zero.
For $\Omega=50$ this behaviour is observed at all scales.
For $\Omega=5$  the skewness drops to a level close to $-0.5$ at the small-scale end, 
a value known from non-rotating flow.
Time-averaged flatness and skewness of the 3D velocity component 
are shown in figure \ref{fig:flatness_skewness}.} 
These findings suggest the existence of a self-similar limiting case for very high
rotation rates \cite[cf.][]{baroud_etal:expspec}.  

{Exemplary one-time PDFs for
the 2D component of the velocity field are shown in figure
\ref{fig:2dpdfs}, the associated time-averaged skewness and flatness functions are displayed in 
figure \ref{fig:flatskew}. In agreement with the weakly intermittent scaling of
the corresponding structure functions, the PDFs at all spatial scales are
close to Gaussian with the flatness increasing for growing $\Omega$ 
from $2.5$ ($\Omega=5)$ to 
$3.1$ ($\Omega=50$) at large scales and decreasing from $3.5$ ($\Omega=5$) to $3.2$ ($\Omega=50$)
at small scales. With increasing rotation rate the skewness also changes from
$-0.1$ (small scales) and about $0$ (large scales) for $\Omega=5$ to positive values close to $0$ at all scales
for $\Omega=50$.
}
}

\section{Turbulent energy decay under rotation}
The influence of rotation on the decay of turbulent energy still
remains a controversial and yet unresolved problem.  To help
clarify this issue the time evolution of kinetic energy in freely
decaying turbulence has been investigated in simulation III for eight
different rotation rates ranging from $\Omega=0$ up to $\Omega = 5$.

Before describing the decaying runs, the macroscopic effects of the sudden onset of rotation on the kinetic energy 
in simulations I and II are reported. In these cases, 
$E$ displays a sharp drop of about $20\%$ (I) and $13\%$
(II) with a subsequent remount that levels off close to the previous
state. 
The dissipation rate $\varepsilon$ also drops but does not
increase again, remaining at $\varepsilon\simeq 0.05$ in both
simulations. This transient behaviour can be understood by the
rotation-induced depletion of the spectral energy transfer that has
been introduced earlier in combination with the applied forcing mechanism. 
For a quasi-stationary state of the
turbulence the time-evolution of the total energy is characterized by
$\dot{E} = \varepsilon_f - \varepsilon = 0$, where $\varepsilon_f$ is
the energy flux injected through the forcing. It is important to note
that the forcing-mechanism used in simulations I and II does not provide a
constant energy injection but rather an $\varepsilon_f$ that nonlinearly adjusts
to the ``demand'' of the fluctuations with $k >k_f$ and is thus coupled to the
cascade dynamics. When rotation is switched on, the cascade gets
depleted and consequently $\varepsilon_f$ diminishes. On the other
hand, as less energy reaches the dissipative scales, $\varepsilon$
also decreases. These two effects lead to a transient evolution of the
energy $E$ until a new equilibrium of forcing, cascade and dissipation
has been reached. The observed behavior does not differ qualitatively
when the rotation is ramped up (as has been checked by test
computations). A similar drop of the initial energy has been reported
by \cite{yeung_zhou:rotdns256}. There, however, the energy did not
remain constant afterwards but increased until the end of the
simulation. This discrepancy is probably due to their stochastic
driving mechanism leading to a pile-up of energy at the largest
scales.  Similarly, the Gaussian forcing spectrum that
\cite{smith_waleffe:rotdns200} applied at intermediate scales led to a
growth of the total energy. 

\begin{figure}
  \centerline{\includegraphics[width=10.0cm]{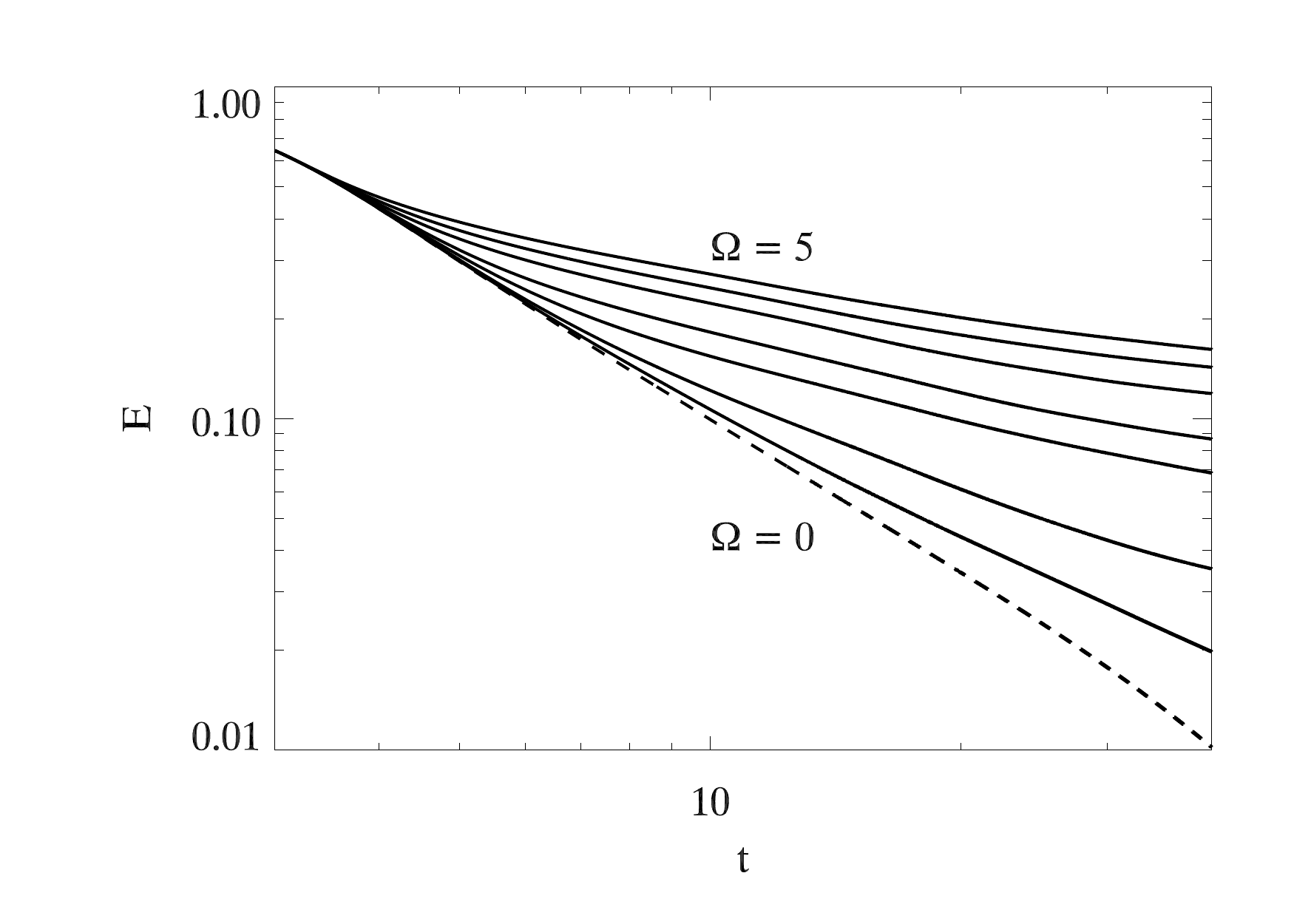}}
    \caption{Decay of the total kinetic energy for \mbox{$\Omega = 0$} (dashed line) and $\Omega = 0.25$, $0.5$, $1$, $1.5$, $2.5$, $3.5$ and $5$ (straight lines from bottom to top).}
    \label{fig:decay_log_3d}
\end{figure}

In the decaying case III rotation is switched on after 1 LET of free evolution, i.e. $t=3$.  
The initial random state with a Gaussian
energy spectrum is similar to that of the driven systems I and II and the same
for all values of $\Omega$. The temporal evolution of $E(t)$ shown in
figure \ref{fig:decay_log_3d} clearly demonstrates that
the energy in rotating turbulence decays slower than in the
non-rotating flow. This effect, a consequence of the attenuation of
the spectral flux, has also been observed experimentally
\cite[see][]{jacquin_etal:exp} and in LES
\cite[see][]{squires_etal:rotles128512,yang_domaradzki:rotles64}. 

\begin{figure}
  \centerline{\includegraphics[width=10.0cm]{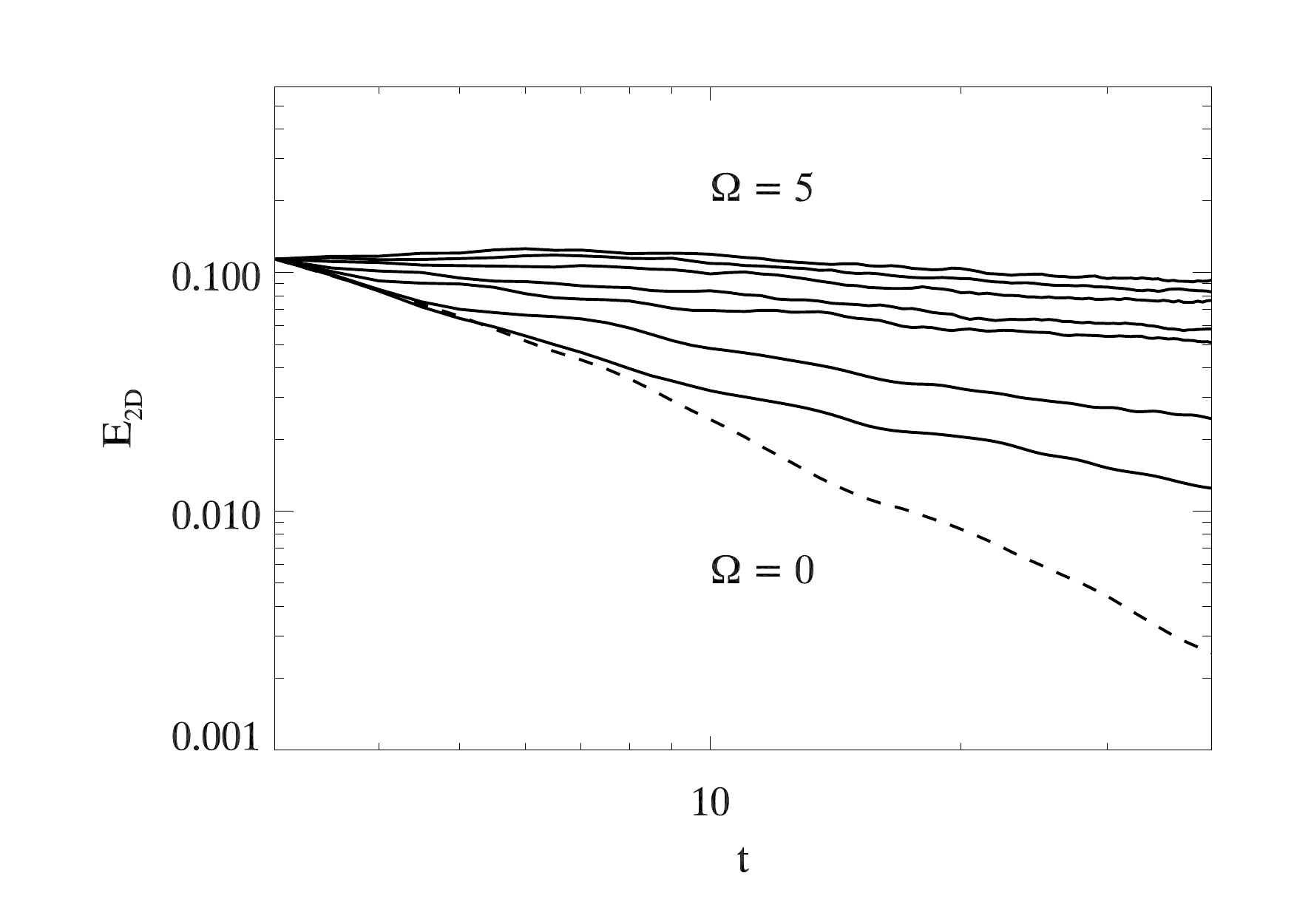}}
    \caption{Decay of the 2D kinetic energy $E_{\mathrm{2D}}(t) = \int d\,k_{\bot}\,E(k_\parallel=0,k_{\bot},t)$ for \mbox{$\Omega = 0$} (dashed line) and $\Omega = 0.25$, $0.5$, $1$, $1.5$, $2.5$, $3.5$ and $5$ (straight lines from bottom to top).}
    \label{fig:decay_log_2d}
\end{figure}

Since the modes with $k_\parallel=0$ play a special role in rotating
turbulence, it is instructive to regard also the 2D kinetic energy
$E_{\mathrm{2D}}(t) = \int
d\,k_{\bot}\,E(k_\parallel=0,k_{\bot},t)$. {As shown in figure
\ref{fig:decay_log_2d}, $E_{\mathrm{2D}}(t)$ tends to be almost constant for
higher rotation rates indicating {a separation in the energetic evolution  
of 2D and 3D
flow components \cite[see, e.g.,][]{bourouiba_bartello:rotdecompo} that becomes significant 
for $\Omega >1.5$. This trend does, however, not lead to total decoupling of 2D and 3D velocity components even in the limit $\Omega\rightarrow\infty$ 
\cite[][]{babin_mahalov_nicolaenko:regularity}.}
Consequently and since the relative contribution of 2D energy compared to 
3D energy in the present simulations is also small, the
decay of total energy in the present simulations is dynamically
governed by the evolution of 3D turbulent fluctuations.  Hence for simplicity and to facilitate 
comparison with other works on energy decay in rotating turbulence, the analysis of
decay properties presented in the following focuses on the (2D+3D) kinetic
energy.}

{
For all rotation rates, the  energy exhibits a period 
of approximate self-similar decay, $E \sim t^{-\alpha}$, see figure \ref{fig:decay_log_3d}.  
In} isotropic Navier-Stokes turbulence this
property has been extensively studied \cite[see
e.\,g.][]{lesieur:book}, and it is of considerable interest to
investigate how it is affected by rotation. During the power-law decay the logarithmic slope 
decreases with increasing rotation rate relating an $\Omega$-dependent decay exponent
$\alpha$. Earlier LES and weak turbulence computations seem to indicate that $\alpha$ is
independent of the rotation rate $\Omega$ with $\alpha \approx
\alpha_0/2$, where $\alpha_0$ is the non-rotating value
\cite[see][]{squires_etal:rotles128512, bellet_etal:weaksimul}. 
Theoretical predictions of
\cite{canuto_dubovikov:formal} and \cite{squires_etal:rotles128512}
based on dimensional and scaling analysis also suggest a scaling
exponent that is constant for all $\Omega$ in contrast to our
numerical results.

\begin{figure}
  \centerline{\includegraphics[width=10cm]{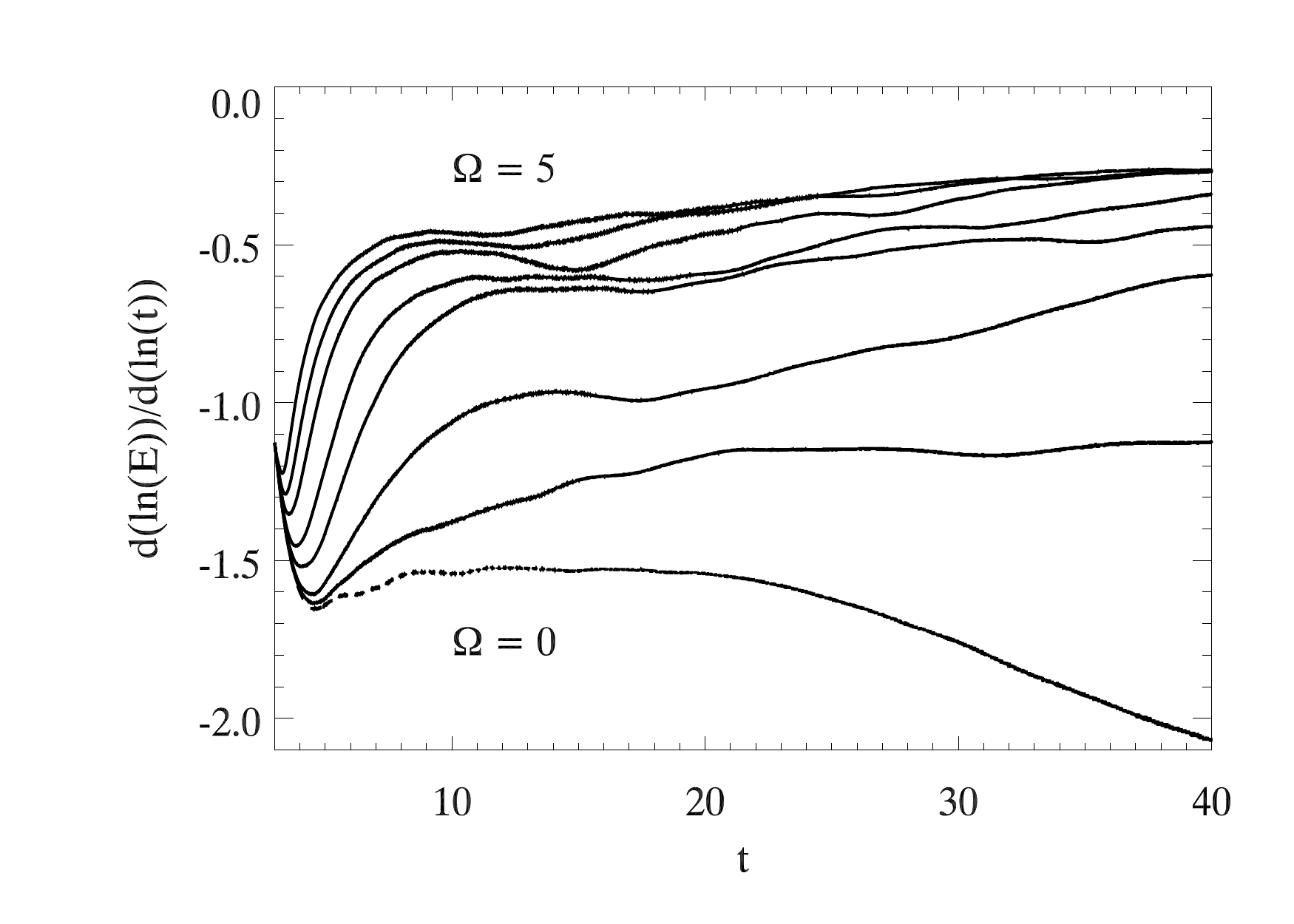}}
    \caption{Logarithmic derivative of the total kinetic energy for \mbox{$\Omega = 0$} (dashed line) and $\Omega = 0.25$, $0.5$, $1$, $1.5$, $2.5$, $3.5$ and $5$ (straight lines from bottom to top).}
    \label{fig:decay_log_deriv_3d}
\end{figure}

\begin{figure}
  \centerline{\includegraphics[width=10cm]{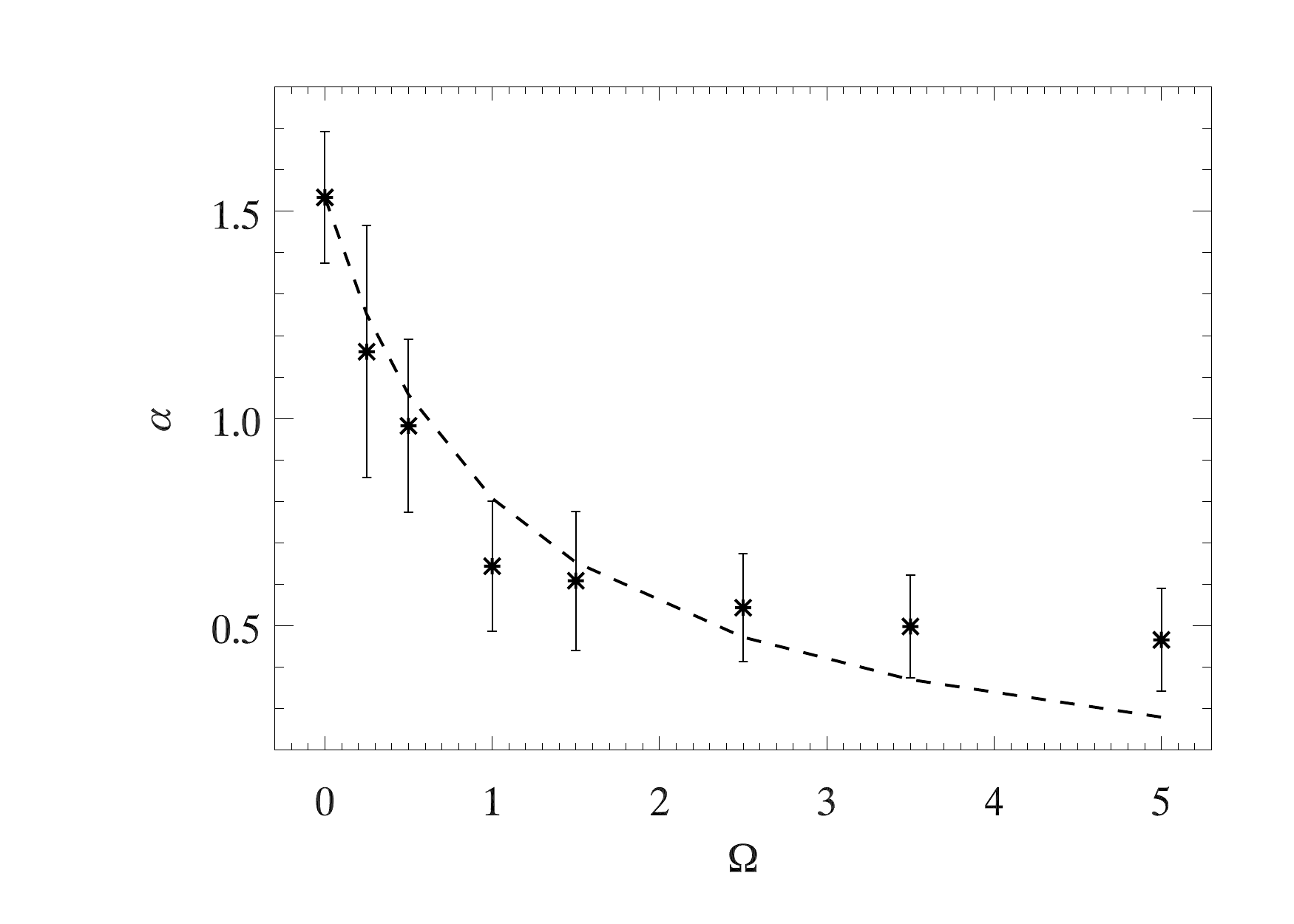}}
    \caption{Decay exponents $\alpha(\Omega)$ for self-similar energy decay in rotating turbulence. The dashed line corresponds to $\alpha(\Omega) = \frac{\alpha_0}{\tau_0'\,\Omega + 1}$ with $\alpha_0 = \alpha(\Omega=0)$ and $\tau_0' \approx 0,9$.}
    \label{fig:decay_exp_3d}
\end{figure}

The actual values of $\alpha_\Omega$ were obtained by determining the
slopes of linear fits to the approximately self-similar regions in the logarithmic
plot of the energy. The respective time intervals were chosen to match the
quasi-constant regions in the logarithmic derivative shown in figure 
\ref{fig:decay_log_deriv_3d} and are listed in
table \ref{tab:decay_intervals}. The different lengths and starting
points of the fit intervals produce some error in the read-off, that
can be estimated by the change of the exponents when determined for
slightly shifted intervals. Finally, we obtain an $\Omega$-dependence
of the decay exponent as shown in figure \ref{fig:decay_exp_3d}. For
$\Omega = 0$ we find $\alpha_0 = 1.54\pm 0.16$ in rough agreement with 
experimental and theoretical decay studies of non-rotating Navier-Stokes turbulence by  
\cite{kolmogorov:k41b}, \cite{saffman:turbdecay}, and \cite{mohsen_larue:expdecay}.
{It should be noted that the absolute values of the 3D energy decay exponents are 
systematically increased by about $0.4$ as compared to the total 3D+2D energy decay while 
exhibiting qualitatively the same $\Omega$-dependence (within error margins).
This is a consequence of the additional quasi-constant contribution of the 2D fluctuations
in the 3D+2D energy.}   

\begin{table}
  \begin{center}
\def~{\hphantom{0}}
  \begin{tabular}{lrrrrrrrr}
$\Omega$	& 0	& 0.25	& 0.5	& 1.0	& 1.5	& 2.5	& 3.5	& 5.0	\\
$t_{min}$	& 11.0	& 16.4	& 10.0	& 11.0	& 9.0	& 7.4	& 7.4	& 6.7	\\
$t_{max}$	& 20.1	& 33.1	& 22.2	& 18.2	& 20.1	& 16.4	& 16.4	& 14.9
\end{tabular}
\caption{Time intervals used for linear fit in figure \ref{fig:decay_log_3d}\,.}
  \label{tab:decay_intervals}
  \end{center}
\end{table}

The dependency of the decay exponent on the rotation frequency can be
explained phenomenologically based on the scaling properties of
rotating turbulence as provided above. First, $E \sim
t^{-\alpha(\Omega)}$ and $\dot{E} \sim \varepsilon$ yields dimensionally
\begin{equation}
	\alpha \frac{E}{\tau_0} \sim \varepsilon\label{decay_law}
\end{equation}
with $\tau_0$ being the large-eddy turnover time. 
Evaluating (\ref{rotscal}) at $L_0$, where most of the total
energy $E$ resides, gives
\begin{equation}
	\varepsilon \sim \frac{E^2}{L_0^2\,\Omega} \sim \frac{E}{\tau_0^2\,\Omega}.\label{decay_epsilon}
\end{equation}
Here, we have used the relation $L_0 \sim E^{\frac{1}{2}}\,\tau_0$. Inserting (\ref{decay_epsilon}) in
(\ref{decay_law}), $\alpha(\Omega)$ is found to follow $\alpha(\Omega)
\sim 1/(\tau_0\,\Omega)$. To include the correct asymptotics for
$\Omega \rightarrow 0$, i.\,e. $\alpha \rightarrow \alpha_0$, we end up
with
\begin{equation}
	\alpha  \sim \frac{\alpha_0^2}{\tau_0\,\Omega + \alpha_0}
\end{equation}
or in a shorter notation,
\begin{equation}
         \alpha  \sim \frac{\alpha_0}{\frac{\tau_0}{\alpha_0}\,\Omega + 1} \sim \frac{\alpha_0}{\tau_0'\,\Omega+1}. \label{decay_alpha}
\end{equation}
Figure \ref{fig:decay_exp_3d} indicates that this simple model
qualitatively reproduces the general features of the
$\Omega$-dependence of the scaling exponent. This $\Omega^{-1}$-dependence suggests a complete depletion
of nonlinear decay in the limit $\Omega \rightarrow \infty$. 
{Such behaviour would also be in accord with a high level of two-dimensionalization, i.e.
$E_\mathrm{2D}\gg E_\mathrm{3D}$.}
We note, however, that the phenomenology presented here
will probably cease to be valid in this asymptotic limit where weak turbulence theory should apply.  
\section{Conclusion}\label{sec:concl}
In summary, high-resolution direct numerical simulations of
large-scale-driven and decaying rotating hydrodynamic turbulence have
been conducted for slow and rapid rotation and analyzed
anisotropically.  With increasing rotation rate, all considered
systems show a trend toward dynamical two-dimensionalization in planes
perpendicular to the rotation axis, strong attenuation of the
nonlinear spectral energy flux along $\boldsymbol{\Omega}$, and
concentration of energy around the plane $k_\parallel=0$.  The
modification of the energy cascade by rotation leads to an energy
spectrum that scales as $\sim k_\perp^{-2}$ in the inertial range with
no discernible scaling in $k_\parallel$. The perpendicular scaling
results from the integration over spectral contributions for all
$k_\parallel$ which exhibit $k^{-3}$-behaviour near $k_\parallel=0$
for low rotation rate.  Also a modified approximate self-similar decay
of the total kinetic energy $E \sim t^{-\alpha(\Omega)}$ is observed
with $\alpha\sim \Omega^{-1}$. {Separating the velocity field
into 2D ($k_\parallel=0$) and 3D ($k_\parallel\neq 0$) components
reveals that the 2D energy tends to become time-independent as the
rotation rate increases.}  {Energy decay and spectral scaling}
can be reproduced by simple phenomenological models which also yield a
non-intermittent scaling relation for fluctuations perpendicular to
the rotation axis, \begin{equation}v_\xi\sim (\varepsilon \Omega)^{1/4}\xi^{1/2}\,,\label{rotscal}
\end{equation}
with $\xi$ being an axis-perpendicular scale. The structure function
scaling and the PDFs {show weak intermittency of the 2D
velocity component while the intermittent signature of the 3D
fluctuations is well represented by a model based on the
She-L\'ev\^{e}que} ansatz. The intermittency of the 3D velocity
component is reflected in the respective PDFs by the well-known change
from Gaussian at large scales to leptocurtic at small scales.  The
two-point statistics show an overall weak dependence on the rotation rate.

\begin{acknowledgments}
The authors thank Stephan K\"ummel, Walter Zimmermann and Lorenz Kramer for their support. 
WCM gratefully acknowledges discussions with J. L\'eorat, A. Mahalov, F. Moisy and J. Seiwert.
\end{acknowledgments}

\bibliographystyle{jfm}

\newcommand{\nop}[1]{}

\end{document}